\documentclass[useAMS,usenatbib]{mn2e}
\usepackage{amsmath}
\usepackage{color}
\usepackage{graphicx}
\usepackage{amssymb}
\usepackage{url}
\usepackage{aas_macros}
\usepackage{booktabs}

\newcommand{\HP}{\mbox{$H_{\rm{P}}\,$}}
\newcommand{\comp}{$\xi_{2.5}$}
\newcommand{\msun}{\ensuremath{\,\textrm{M}_{\odot}}}
\newcommand{\proto}{\ensuremath{M_{\textrm{proto}}}}
\newcommand{\mco}{\ensuremath{M_{\textrm{CO}}}}
\newcommand{\mfe}{\ensuremath{M_{\textrm{Fe}}}}
\newcommand{\mrem}{\ensuremath{M_{\textrm{rem}}}}
\newcommand{\mzams}{\ensuremath{M_{\textrm{ZAMS}}}}

\newcommand{\mfin}{\ensuremath{M_{\textrm{fin}}}}
\newcommand{\ffb}{\ensuremath{f_{\textrm{fb}}}}
\newcommand{\mfb}{\ensuremath{M_{\textrm{fb}}}}

\newcommand{\starlab}{\textsc{Starlab}}
\newcommand{\starlabmm}{\textsc{StarlabMM}}
\newcommand{\kira}{\textsc{Kira}}
\newcommand{\seba}{\textsc{SeBa}}
\newcommand{\startrack}{\textsc{StarTrack}}
\newcommand{\sse}{\textsc{SSE}}
\newcommand{\parsec}{\textsc{PARSEC}}
\newcommand{\nbody}{\ensuremath{N}-Body}
\newcommand{\nbodyx}{\textsc{NBODYx}}
\newcommand{\higpus}{\textsc{HiGPUs}}
\newcommand{\mesa}{\textsc{MESA}}
\newcommand{\amuse}{\textsc{AMUSE}}
\newcommand{\codename}{\textsc{SEVN}}

\title[The mass spectrum of compact remnants from the \textsc{PARSEC} stellar evolution tracks]{The mass spectrum of compact remnants from the \textsc{PARSEC} stellar evolution tracks}
\author[Spera et al.]{Mario Spera$^{1}$\thanks{E-mail: mario.spera@oapd.inaf.it or mario.spera@live.it}, 
Michela Mapelli$^{1,2}$, Alessandro Bressan$^{3,1}$\\
$^{1}$INAF, Osservatorio Astronomico di Padova, Vicolo dell'Osservatorio 5, I-35122, Padova, Italy\\
$^{2}$INFN, Milano Bicocca, Piazza della Scienza 3, I--20126, Milano, Italy\\ 
$^{3}$ Scuola Internazionale Superiore di Studi Avanzati (SISSA), Via Bonomea 265, I-34136, Trieste, Italy}

\begin{document}

\maketitle

\begin{abstract}
The mass spectrum of stellar-mass black holes (BHs) is highly uncertain. Dynamical mass measurements are available only for few ($\sim{}10$) BHs in X-ray binaries, while theoretical models strongly depend on the hydrodynamics of supernova (SN) explosions and on the evolution of massive stars. In this paper, we present and discuss 
the mass spectrum of compact remnants that we obtained with \codename{}, a new public population-synthesis 
code, which couples  the \parsec{} stellar evolution tracks with up-to-date recipes for SN explosion 
(depending on the Carbon-Oxygen mass of the progenitor, on the compactness of the stellar core at pre-SN 
stage, and on a recent two-parameter criterion based on the dimensionless entropy per nucleon at pre-SN 
stage). \codename{} can be used both as a stand-alone code and in combination with direct-summation N-body 
codes 
(\starlab{}, \higpus{}).  The PARSEC stellar evolution tracks currently implemented in SEVN predict significantly larger values of the Carbon-Oxygen core mass with respect to previous models. For most of the SN recipes we adopt, this implies substantially larger BH masses at low metallicity ($\leq{}2\times{}10^{-3}$), than other population-synthesis codes. The maximum BH mass found with \codename{} is $\sim{}25$, 60 and 130 \msun{} at metallicity $Z =2 \times{} 10^{-2}$ , $2 \times{}10^{-3}$ and $2\times{} 10^{-4}$ , respectively. Mass loss by stellar winds plays a major role in determining the mass of BHs for very massive stars 
($\geq{}90$ M$_\odot{}$), while the remnant mass spectrum depends mostly on the adopted SN recipe for lower progenitor masses. We 
discuss the implications of our results for the transition between NS and BH mass, and for the 
 expected number of massive BHs (with mass $>25$ M$_\odot{}$) as a function of metallicity.

\end{abstract}


\section{Introduction}
Compact remnants are the final stage of the evolution of massive stars, and power a plethora of important
astrophysical processes: they are the engine of the X-ray binaries we observe in the nearby Universe,  and
may be powerful sources of gravitational waves  (e.g. \citealt{phinney1991}). Furthermore, the merger of two 
neutron stars
(NSs) and/or that of a stellar black hole (BH) with a NS are expected to lead to one of the most energetic transient
phenomena in the Universe: the short gamma-ray bursts (e.g. \citealt{paczynski1991}). Finally, X-ray binaries
powered by BHs and/or NSs are the key to explain some of the most luminous point-like non-nuclear X-ray 
sources (the
ultraluminous X-ray sources, e.g. \citealt{mapelli2010,mapelli2014} and references therein), and are an
important source of feedback, in both the nearby and the early Universe (e.g. \citealt{justham2012}, and
references therein).

Despite their importance for astrophysics, the details of the formation of BHs and NSs (and especially the
link with their progenitor stars) are matter of debate. From the observational point of view, the confirmed
BHs are only a few tens (see table 2 of \citealt{ozel2010}, for one of the most updated compilations). These
are located in X-ray binaries, mostly in the Milky Way (MW), and an accurate dynamical mass estimate has been
derived only for a fraction of them ($\sim{}10$). Most of the derived BH masses are in the range
$5\le{}m_{\rm BH}/\msun\le{}10$. In the MW, the most massive BHs in X-ray binaries do not
significantly exceed $m_{\rm BH}\sim{}15\msun$, whereas a few BHs in nearby galaxies might have
higher masses: M33 X-7 ($m_{\rm BH}=15.65\pm{}1.45\msun$, \citealt{orosz2007}), IC-10 X-1
($m_{\rm BH}\sim{}23-34\msun$, \citealt{prestwich2007, silverman2008}), NGC 300 X-1 ($m_{\rm BH}>10\msun$,
\citealt{crowther2007, crowther2010}). Interestingly, these three massive BHs are in
regions with relatively low metallicity. A metallicity $Z\sim{}0.004$ is estimated for the dwarf irregular
galaxy IC-10 \citep{garnett1990}. The metallicity of M33 in proximity of  X-7  is $Z\sim{}0.008$, and that of
NGC300 in proximity of X-1 is $Z\sim{}0.006$  \citep{pilyugin2004}.

The statistics is significantly larger for NSs: currently, there are dynamical mass measurements for 61 NSs (17, 11, 30, and 3 of them are in X-ray binaries, NS-NS binaries, NS-white dwarf binaries and NS-main sequence binaries,
respectively, {\tt http://stellarcollapse.org/nsmasses}, \citealt{lattimer2005,lattimer2012}).

The link between the progenitor star and the compact remnant is still poorly constrained
for both BHs and  NSs: observations of core-collapse supernovae (SNe) indicate a deficit of massive
($\gtrsim{}20\msun$) progenitor stars
(\citealt{smartt2009,horiuchi2011,jennings2012,jennings2014,gerke2014}), which possibly suggests that  the
most massive stars undergo no or faint SNe.

From a theoretical perspective, the formation and the mass spectrum of BHs and NSs strongly depend on two
fundamental processes: (i) the hydrodynamics of SNe; (ii) mass loss by stellar
winds in massive stars (during and especially after the main sequence, MS).

 (i) The physics of SN explosions is extremely complex, and the hydrodynamical codes that investigate the 
explosion mechanisms are computationally challenging (see e.g. 
\citealt{fryer1999,fryer2001,heger2002,heger2003,fryer2006,oconnor2011,fryer2012,janka2012,ugliano2012,burrows2013,pejcha2015,ertl2015}).
 In particular, the link between the late evolutionary stages of a massive star and the SN products is still 
matter of debate. Several authors (e.g. \citealt{bethe1990,janka2007,burrows2013,janka2012}) investigate for 
which structural properties of the progenitor star a SN can fail, leading to the direct collapse of the star 
to a BH. Even if the SN occurs, how much matter can fall back and be accreted onto the proto-compact remnant 
is very uncertain.

(ii) For massive progenitors (zero-age MS mass $\mzams\ge{}30 \msun$) the details of stellar
evolution are  very important for the SN outcome and for the final remnant mass. In fact, the
 final mass \mfin{} of the progenitor star (i.e. the mass of a star immediately before the collapse) is 
 governed by the amount of mass loss by stellar winds (e.g.
\citealt{mapelli2009, belc2010, fryer2012,mapelli2013}).
The rate of mass loss by stellar winds on the MS increases with the metallicity of the star as
$\dot{M}\propto{}Z^{\alpha{}}$, where $\alpha{}\sim{}0.5-0.9$, depending on the model (e.g.,
\citealt{kudritzki1987, leitherer1992, kudritzki2000, vink2001, kudritzki2002}). The behaviour of evolved
massive stars, such as luminous blue variable stars (LBVs) and Wolf-Rayet stars (WRs), is also expected to
depend on metallicity, but with larger uncertainties (e.g., \citealt{vink2005,meynet2005,bressan2012,tang2014}).

Both the models of SN explosion  (e.g. \citealt{fryer2012,janka2012,burrows2013,pejcha2015,ertl2015}) and the theory of massive star evolution   (e.g. \citealt{bressan2012,tang2014}) were deeply revised in the last few 
years. For these reasons, population synthesis 
codes that aim at studying 
the demographics of compact remnants must account for up-to-date models for both SN explosions and stellar 
evolution. Here we present \codename{}  (acronym for `Stellar EVolution N-body'), a new population synthesis tool that couples \parsec{} evolutionary tracks for stellar evolution \citep{bressan2012, chen2014, tang2014}
with  up-to-date models for SN
explosion (\citealt{fryer2012,janka2012,ertl2015}), and that can be  easily merged with several N-body codes.
 The new \parsec{} evolutionary tracks  consider the most recent updates
for mass loss by stellar winds and other input physics. In
this paper, we present and discuss the mass spectrum of BHs and NSs that we obtain from \codename{}, with
particular attention to the dependence of the remnant mass on metallicity. 

Furthermore, \codename{} is extremely versatile, because it relies upon a set of tables extracted
from stellar evolution tracks: if we are interested in comparing different 
stellar evolution models, we can
do it quickly and easily, by changing tables. The new tool is publicly available\footnote{ \codename{} upon
request to the authors, through the email \texttt {mario.spera@oapd.inaf.it} or 
\texttt{mario.spera@live.it}}.  
 \codename{} is specifically designed to add updated recipes for stellar evolution and SN explosion to \nbody{} simulations, even though it can be used as a simple and fast stand-alone 
population-synthesis code too. In particular, we  merged it with the \starlab{} public 
software environment \citep{zwart2001}
and with an upgraded version of \higpus{} code (\cite{dolcetta2013}; Spera, in preparation).  Thus, the new 
code can be used for both population
synthesis studies of compact-object binaries in the field, and for investigating the dynamical evolution of compact objects in star clusters.   The evolution of compact remnants in star clusters is of crucial importance, since star
clusters are sites of intense dynamical processes, which may significantly affect the formation of X-ray
binaries (e.g. \citealt{blecha2006, mapelli2013,mapelli2014}), as well as the formation and merger of
double-compact object binaries (e.g. \citealt{oleary2006,sadowski2008,downing2010,downing2011,ziosi2014}).
Furthermore, extreme dynamical processes, such as repeated mergers of compact remnants (\citealt{miller2002})
and the runaway merger of massive objects in star clusters (\citealt{zwart2002}), can lead to the formation
of intermediate-mass BHs (i.e. BHs with mass $10^2-10^5$ M$_\odot$). Finally, compact remnants are also 
expected to affect the overall dynamical evolution of star clusters 
\citep{downing2012,sippel2012,mapelli2013b,trani2014}. 

This paper is organized as follows. In Section~\ref{sec:method}, we describe the main features and 
ingredients of \codename{} (including stellar evolution and SN models). In Section~\ref{sec:results}, we 
discuss  the outputs of \codename{}, with particular attention to the mass spectrum and the mass function of 
NSs and BHs. Furthermore, we compare the results of \codename{} with those of other population-synthesis 
codes. In Section~\ref{sec:newmodels}, we discuss  the results we obtained applying the \citet{oconnor2011} and \citet{ertl2015} prescriptions for SN explosion to \parsec{} progenitors, at metallicity $Z=0.02$. In 
Section~\ref{sec:conclusions}, we summarize our main results.

\section{Method}\label{sec:method}

\subsection{Single stellar evolution with \parsec{}}
\label{subsec:parsec_description}
The \parsec{} database  includes updated and homogeneous
sets of canonical single stellar evolutionary tracks,
from very low ($M$=0.1\msun) to very massive ($M$=350\msun) stars, and
from the pre-MS to the beginning of central carbon burning. 
The code is thoroughly discussed  in \citet{bressan2012, bressan2013}, \citet{chen2014} and \citet{tang2014} and
here we briefly describe its most important characteristics.
The equation of state (EOS) is computed with the FreeEOS code\footnote{http://freeeos.sourceforge.net/}
(A.W.~Irwin).
Opacities are computed combining the high-temperature data from the Opacity Project At Lawrence Livermore National Laboratory (OPAL)
\citep{IglesiasRogers_96} with the low-temperature data from the 
\AE SOPUS\footnote{http://stev.oapd.inaf.it/aesopus} code \citep{MarigoAringer_09}.
Conductive opacities are included following \citet{Itoh_etal08}.
The main Hydrogen and Helium burning reactions are included as recommended in the JINA  database
\citep{Cyburt_etal10} with electron screening factors taken from
\citet{Dewitt_etal73} and \citet{Graboske_etal73}. Energy losses by electron neutrinos are taken from
\citet{Munakata_etal85} and \citet{ItohKohyama_83} and \citet{Haft_etal94}.
Instability against convection is tested by means of the Schwarzschild criterion
and, where needed, the convective temperature gradient is estimated with the mixing-length theory of \citet{mlt}
with a mixing length parameter calibrated on the solar model, $\alpha_{\rm MLT}=1.74$.
The location of the boundary of the convective core is estimated
in the framework of the mixing-length theory, allowing for the penetration of convective
elements into the stable regions \citet{bressan1981}.
As thoroughly described in \citet{bressan2013}, 
the main parameter describing core overshooting
is the mean free path of convective elements {\em across} the border of the unstable region
$l_{\rm c}$=$\Lambda_{\rm c}$\HP with
$\Lambda_{\rm c}=0.5$, as result of the calibration obtained by the analysis of intermediate age clusters \citep{Girardi_etal09}
as well as individual stars \citep{Kamath_etal10, Deheuvels_etal10, Torres_etal14}.
Effects of stellar rotation have not yet been introduced in \parsec{}.

The reference solar partition of heavy elements is taken from \citet{Caffau_etal11} who revised
a few species of the \citet{GrevesseSauval_98} compilation.
According to \citet{Caffau_etal11} compilation,  the present-day Sun's
metallicity is $Z_{\odot}= 0.01524$.

While the evolution below $M=12\msun$ is computed at constant mass, for more massive stars the mass loss rate
is taken into account combining  the mass-loss rates formulations provided by different authors
for different evolutionary phases, as described in \cite{tang2014}.
During the  Blue Super Giant (BSG) and LBV  phases
we adopt the maximum between the relations provided by \citet{Vink_etal00,Vink_etal01},
and that provided by \citet{Vink_etal11} which includes
the dependence of the mass-loss rates on the ratio  ($\Gamma$) of the star luminosity to the corresponding 
Eddington luminosity.  In the Red Supergiant (RSG) phases we adopt the mass-loss rates by 
\citet{de_Jager_etal88}, R$_{dJ}$, 
while, in the  WR phases, we use the \citet{Nugis_etal00} formalism.

An important effect of the metallicity is its modulation of the mass loss rates. As discussed in 
\citet{tang2014} and in Chen et al. (2015, in preparation), the dependence of the radiation driven mass-loss 
rates on the metallicity is a strong function of $\Gamma$. While, at low values of $\Gamma$, the mass-loss 
rates obey the relation $\dot{M}\propto(Z/Z_G)^{0.85}\,{\rm M_\odot{}}$ yr$^{-1}$ 
\citep{Vink_etal00,Vink_etal01}, with $Z_G=0.02$ being the average metallicity assumed for Galactic massive 
stars, at increasing $\Gamma$ the metallicity dependence becomes weaker, and it disappears as $\Gamma$ 
approaches 1 \citep{Grafener_etal08}. \citet{tang2014} show that the metallicity effect can be expressed as
\begin{equation}
\dot{M}\propto~(Z/Z_G)^\alpha,
\label{mdotgamma}
\end{equation}

with the coefficient $\alpha$ determined from a fit to the published relationships by \citet{Grafener_etal08}
\begin{eqnarray}
\alpha =& 0.85 \qquad \qquad  & (\Gamma < 2/3) \nonumber\\
\alpha =& 2.45-2.4\,{}\Gamma \qquad \qquad & (2/3\leq\Gamma\leq{1})
\label{mdotalpha}
\end{eqnarray}

In the  WR phases, \parsec{} makes use of the \citet{Nugis_etal00} formalism, with its own dependence on the 
stellar metallicity while, during the Red Supergiant (RSG) phases the \citet{de_Jager_etal88} rates are 
re-scaled  adopting the usual relation $\dot{M}\propto(Z/Z_G)^{0.85}\,{\rm M_\odot}$ yr$^{-1}$.

With these assumptions for the mass-loss rates, the new models of near-solar metallicity can naturally reproduce the observed lack of supergiant stars above the \citet{Humphreys_Davidson_1979} limit.
The lack of RSG stars is usually interpreted as a signature of the
effects of enhanced  mass-loss rates when the star enter this region,
and this interpretation is supported by the presence, around this limit,
of LBV stars which are known to be characterized by high mass loss rates.
While, in previous models, the limit was reproduced by adopting an ``ad-hoc'' enhancement of the mass-loss rates,
in the current models  the enhancement is nicely reproduced by
the boosting of the mass-loss rate when the stars approach the Eddington limit (Chen et al., in preparation).
At metallicities lower than solar,  the boosting is mitigated by the reduction factor introduced by the metallicity dependence.
At $Z=0.001$, the upper MS widens significantly  and the more massive stars 
evolve in the ``forbidden''  region even during the H-burning phase, because of their very large convective 
cores.
They may also ignite and burn central helium as ``red'' super-giant stars. The full set of new 
evolutionary  tracks and the corresponding isochrones may be found at 
\url{http://people.sissa.it/~sbressan/parsec.html} and \url{http://stev.oapd.inaf.it/cgi-bin/cmd}, 
respectively.

\subsection{\codename{} general description}
\label{subsec:stevoimpl}
The coupling between dynamics and stellar evolution, in a single code, can be achieved through 
 three alternative approaches:
\begin{itemize}
\item the first one is based on a ``brute force'' approach. It consists in calling an advanced stellar 
evolution code (such as \parsec{}) that calculates the detailed evolution of stellar physical parameters step 
by step, following the time intervals imposed by the \nbody{} dynamics;
\item the  second one is based on polynomial fittings that interpolate the fundamental 
stellar parameters
(radius, luminosity, temperature and chemical composition), as a function of time, mass and metallicity.
Besides being a fast choice in terms of computing
time, one of the main advantages of using this strategy is that it can be implemented with little effort;
\item the  third approach consists in using stellar evolution isochrones as input files. 
These isochrones are usually
provided in the form of tables, for a grid of masses and metallicities, and they are read and interpolated
by the numerical code on the fly. The main advantage of this strategy is that it makes the implementation
more general. The option to change the built-in stellar evolution recipes is left to users, who can
substitute the input tables, without modifying the internal structure of the code or even recompiling it.
\end{itemize}

The first approach is highly inefficient because the continuous calls to advanced stellar evolution 
codes, inside an \nbody{} integrator, significantly slows down the overall numerical evolution. To 
develop \codename{}, we chose to follow the second aforementioned approach
(usage
of stellar evolution isochrones in tabular form).
 \codename{} can work as a stand-alone code (for fast population synthesis studies in the field), and can be 
 linked to a large variety of \nbody{} codes,  without suffering a performance penalty. In 
 particular, we merged \codename{} with an updated version of the direct \nbody{} code
\higpus{}\footnote{\url{http://astrowww.phys.uniroma1.it/dolcetta/HPCcodes/HiGPUs.html}}
(\citealt{dolcetta2013}; Spera, in preparation) as well as in the \starlab{} software
environment\footnote{\url{http://www.sns.ias.edu/~starlab/}} \citep{zwart2001}, and it can also be included
in the Astrophysical
Multipurpose Software Environment (\amuse{}\footnote{\url{http://amusecode.org/wiki}},
\citealt{pelupessy2013}).

In this paper, we focus our attention on our implementation of \codename{} in \starlab{} since it already 
includes both an \nbody{} integrator (called \kira) and a binary evolution module (\seba{}).
In particular, we updated a version of \seba{} that had been previously modified by \citet{mapelli2013}, who included
metallicity dependent stellar winds \citep{hurley2000} and prescriptions for the mass loss
by MS stars \citep{vink2001}. While we left the dynamical integration part untouched, we
rearranged \seba{} by adding stellar isochrone tables, at different metallicity, and by forcing  the
software to use them as input files. In this way, we have hidden the default implementation without making
radical changes to the code structure. In the current version of \codename{},  we use the \parsec{} data 
to get the physical parameters of the stars for
all evolutionary stages but the thermally-pulsating AGB phase (TP-AGB). In fact,
the evolution and lifetimes of TP-AGB stars suffer from significant uncertainties and a thorough calibration 
of the
latter phase is still underway \citep{marigo2013,rosenfield2014}.
At present, we use
the built-in \seba{} \texttt{super\_giant} class to follow the evolution of the stars in this stage.
Moreover, according to the \parsec{} recipes, all stars with an initial mass
$\mzams \lesssim  M_{up}$ (with $M_{up} = 7 \msun$) undergo the AGB phase. In particular, at the end of
their lives, stars of mass
$\mzams \gtrsim M_{up}$ will explode as
SNe leaving NSs or BHs as compact remnants, while stars with $\mzams <
M_{up}$ will
evolve through the AGB phase, quickly losing their envelopes, until a WD is formed. More technical details 
about the \codename{} implementation can be 
found in Appendix~\ref{appsec:detailsimplementation}.

\subsection{Prescriptions for the formation of compact remnants}

The default recipes implemented in the \seba{} module predict the formation of a white dwarf (WD) if the 
final core mass is less than the Chandrasekhar mass ($1.4\msun$), a NS or a BH  if the core
mass is greater than $1.4\msun$. In our implementation of \codename{} in \seba{}, we leave the
recipes for the formation of WDs unchanged, but we change the way to form NSs and BHs. 

The default version of \seba{} distinguishes between NSs and BHs by inspecting the final mass of the
core: if it is larger than the Chandrasekhar mass ($1.4 \msun$) and,
at the same time, the initial mass of the star is $\mzams < 25\msun$, a NS is formed. If $\mzams \geq
25\msun$ or if the final carbon-oxygen (CO) core mass ($\mco$) is such that $\mco \geq  5\msun$, the star
ends its life forming a BH\footnote{In \seba, the limits $25\msun$ and $5\msun$  are the default values of
two parameters called \texttt{super\_giant2black\_hole} and
\texttt{COcore2black\_hole}, respectively. The user can adjust them at choice.}.  To determine the BH
mass, \seba{} assumes that, initially, a fixed amount of the CO core mass collapses, forming a proto-compact
object of mass $\proto=3\msun$ \citep{fryer2001}. The amount of fallback material, $\mfb$, is determined by
comparing the binding
energies of the hydrogen (H), helium (He) and CO shells with the SN explosion energy.  
 The final mass of the compact object is
given by $M_{\textrm{BH}}=\proto+\mfb$.

 In \codename{}, we  
substituted the default {\sc seba} treatment of SNe with the following new recipes.
 We implemented the three models described in details by \citet{fryer2012}: (i) the model implemented in the \startrack{} population synthesis code
(see \citealt{belc2008,belc2010}), (ii) the
\textit{rapid supernova
model}, and (iii) the \textit{delayed supernova model}. The main difference between the last two
explosion mechanisms is the time-scale over which the explosion occurs: $<250$ ms after the bounce for
the rapid model, $\gtrsim 0.5$ s for the delayed mechanism (for the details see, for example, \citealt{bethe1990}). A common feature of  these models is that they depend only on the final characteristics of
the star, by means of the final CO core mass (\mco{}) and of the final mass of the star (\mfin{}).  Appendix ~\ref{appsec:summaryfryer} 
summarizes the main features of the \citet{fryer2012} SN explosion recipes.  
 We recall that the \cite{fryer2012}
methods are general prescriptions for the formation of
compact remnants, and do not distinguish, {\it a priori}, between NSs and BHs.  In \codename{}, we assume that all the remnants with masses $\mrem <
3.0\msun$ are NSs, and that the objects with masses $\mrem \geq 3.0\msun$ are BHs, according to the maximum
mass of a NS indicated by the Tolman-Oppenheimer-Volkoff limit \citep{oppen1939}. While the \citet{fryer2012} 
models are extremely simple to implement in a population-synthesis code, it has been recently suggested that 
the dependence of the mass of the compact remnant on \mfin{} or \mco{} might be significantly more complex 
(\citealt{ugliano2012,oconnor2011,sukhbold2014,janka2012,smartt2015}). The internal structure of stars, at 
core-collapse stage, may exhibit significant differences, leading to deep changes on the physical parameters 
of compact remnants, even if the progenitors are very close in terms of \mzams{} or \mco{}. As a 
consequence, a one to one relation between the mass of the compact remnant and, e.g., \mco{} could be  
inadequate to discriminate between SNe (formation of a NS) and failed SNe (direct collapse to a BH). The 
critical parameter to distinguish between SNe and failed SNe might the compactness of 
stellar cores at the pre-SN stage (\citealt{oconnor2011,ugliano2012,sukhbold2014}). Alternatively one may use 
an equivalent criterion based on the two-parameters $M_4$, representing the enclosed mass at a dimensionless 
entropy per nucleon $s = 4$, and $\mu{}_4$, that is the mass gradient at the same location \citep{ertl2015}. 
In order to fulfil these recent advances of the SN explosion models and to test their impact on the mass 
spectrum of compact remnants, we have implemented in \codename{} these  two additional SN explosion recipes,  
 namely the criterion based on the compactness of stellar cores \citep{oconnor2011,ugliano2012,sukhbold2014} 
 and the criterion  
based on $M_4$ and $\mu{}_4$ \citep{ertl2015}. We present here only the results for 
$Z=0.02$, because the results at metallicities lower than solar are still under investigation.

In \codename{}, we set the
delayed SN model as default SN explosion mechanism, but the user can
choose one of the  aforementioned mechanisms by modifying the input parameter file. Only for the  
\codename{} implementation in \starlab{}, we also leave the choice to use the \seba{} built-in 
recipes\footnote{The first line of the
file \texttt{input\_param.txt} determines the SN explosion model that will be adopted throughout the
numerical simulation. It can be \texttt{delayed}, \texttt{startrack}, \texttt{rapid}, \texttt{compactness} or \texttt{twoparameters}. In the implementation 
of \codename{} in \starlab{}, this line can be even \texttt{default}  if we want to use the \seba{} built-in 
recipes for SN explosion.}.

Furthermore, the aforementioned models do not account for the possibility that the progenitor undergoes a pair-instability SN 
(e.g. \citealt{woosley2002}). In \codename{}, we add the option to activate pair-instability SNe, when the 
Helium core mass (after the He core burning phase) is $60\leq{}M_{\rm He}/\msun{}\leq{}133$. For this range 
of He core masses, the star does not leave any remnant, while it directly collapses to BH for larger masses. 
In the following Section, we show models that do not  undergo
pair-instability SNe. 

When a compact remnant is formed, it also receives a velocity kick, $W_{kick}$, due to the
asymmetries that can occur during the collapse process. In \codename{}, we determine the absolute value of
the kick using the three dimensional velocity distribution of
the pulsars observed in our galaxy. For details, we refer to \citet{hobbs2005}, who studied the proper
motions of $233$ pulsars, obtaining a Maxwellian fit for their velocity
distribution, with a one dimensional variance equal to $256$ km/s. The direction of the kick is randomly
chosen. 
Furthermore,
following the prescriptions given in \citet{fryer2012}, we also included the
dependence of the velocity kick on the amount of mass that falls back onto the proto-compact object.
Specifically, the actual value of the
kick imparted to a compact remnant, $V_{\rm kick}$, is given by
\begin{equation}
V_{kick}=\left(1-\ffb\right)W_{kick}.
\end{equation}
Thus, a BH that forms via direct collapse ($\ffb=1$) does not receive a velocity kick, while
full kicks are assigned to compact remnants formed with no fallback. Another possible treatment for BH kick velocities, is to assume that BHs follow the same distribution of $W_{\rm kick}$ as NSs, but normalized to $\langle{}M_{\rm NS}{}\rangle{}/ M_{\rm BH}$ (where $\langle{}M_{\rm NS}\rangle{}$ is the average NS mass), to ensure momentum conservation. We leave this second option in \starlab{}, even if we set the former treatment as default.

\section{Results}\label{sec:results}

In this section, we discuss the effects of metallicity on the stellar mass loss rate, on the CO core, on the 
formation of
compact remnants, and on the mass function of NSs and BHs, as we found using our new tool \codename{}. 
We also discuss the main differences between \codename{} and other population synthesis codes, in terms of mass spectrum of compact remnants. In particular, we compare the results  of
\codename{} with those of \sse{} \citep{hurley2000},
of \starlab{} \citep{zwart2001}, and of the version of \starlab{} modified by \citet{mapelli2013} (hereafter referred as \starlabmm{}). In particular,
\sse{} is a stellar evolution tool that has already been linked to the  \nbodyx{} family of \nbody{} codes (see 
e.g. \citet{aarseth1999} and \citet{nitadori2012}) and it also implements recipes for metallicity-dependent 
stellar winds. Moreover, \sse{} adopts the SN explosion recipes described in \citet{belc2002}.

\subsection{Mass loss by stellar winds}

\begin{figure}
\centering
\includegraphics[scale=0.37]{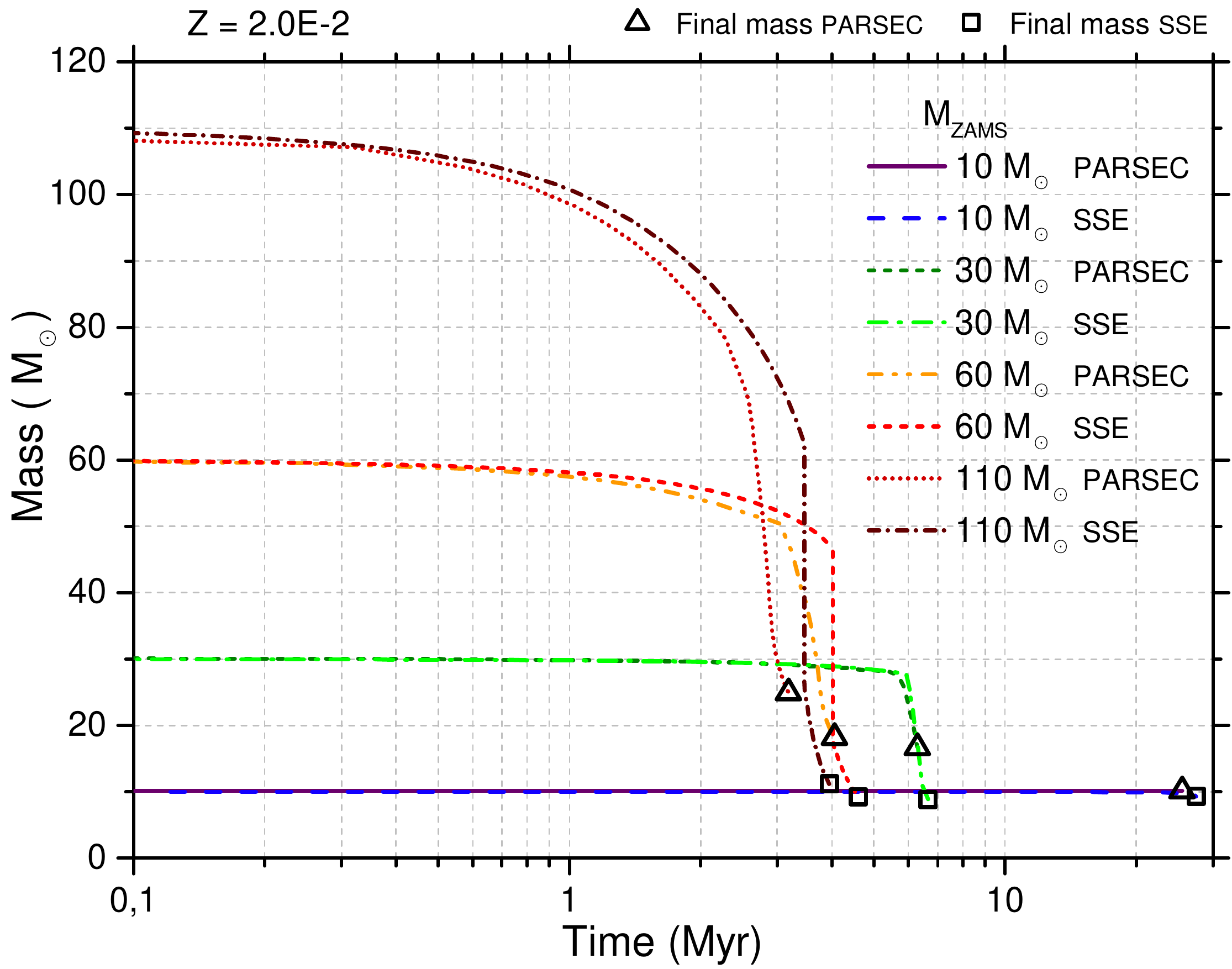}
\caption{Temporal evolution of the stellar mass for different \mzams{} and for metallicity 
$Z=2.0\times{10^{-2}}$,
according to \parsec{} and \sse{}.
Solid (dashed) line: evolution of a star with $\mzams=10\msun$ with \parsec{} (\sse{}); dotted (dash-dotted) line: evolution of a star with $\mzams=30\msun$ with \parsec{} (\sse{}); dash-double dotted (short dashed) line: evolution of the mass of a star
with $\mzams=60\msun$ with \parsec{} (\sse{}); short dotted (short dash-dotted) line: evolution of the mass of a star with $\mzams=110\msun$ with \parsec{} (\sse{}). Open triangles and open squares mark the final
point of each curve obtained using \parsec{} and \sse{}, respectively.}
\label{fig:fig1}
\end{figure}

\begin{figure}
\centering
\includegraphics[scale=0.37]{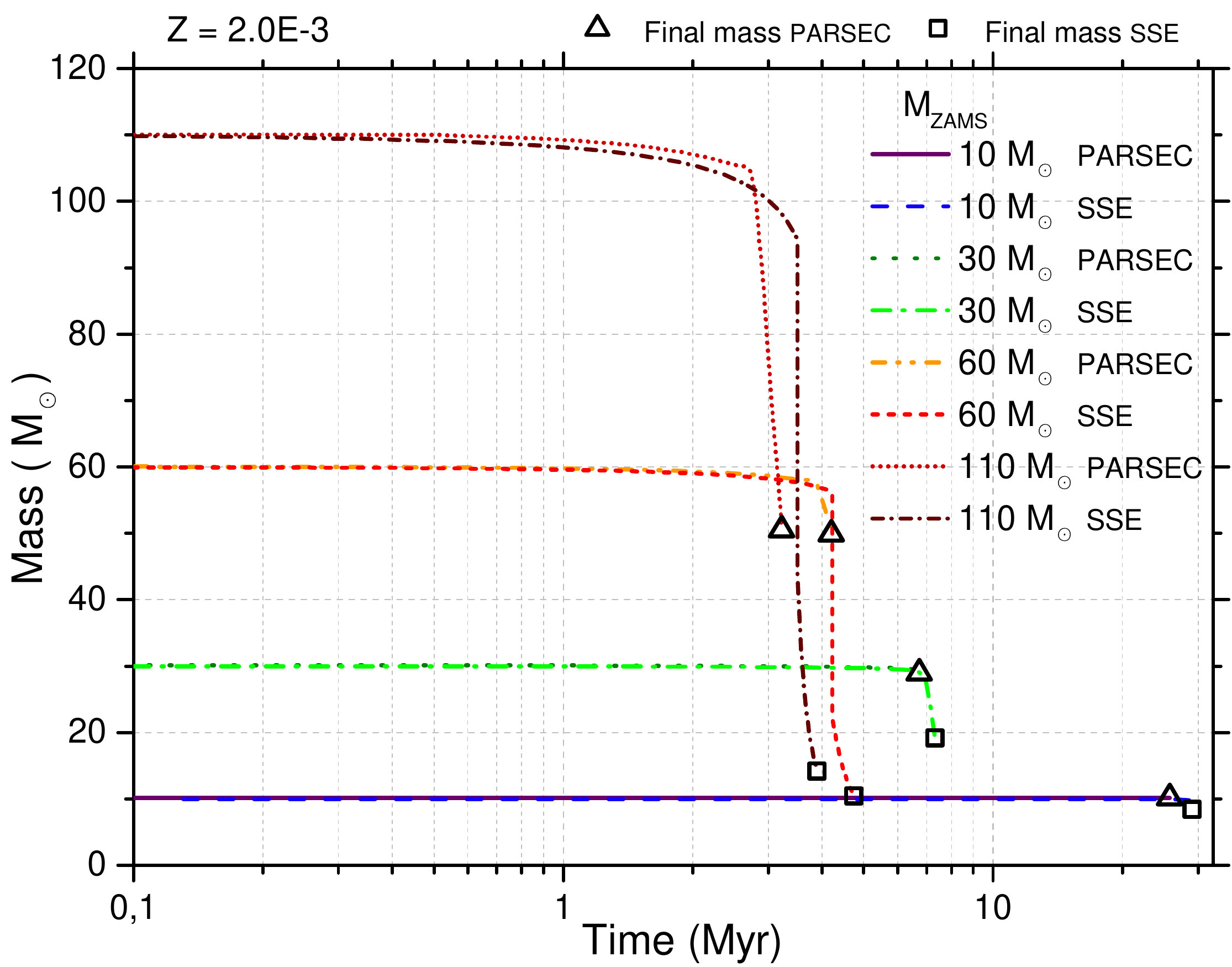}
\caption{Same as Fig. \ref{fig:fig1}, but for $Z=2.0\times{10^{-3}}$.}
\label{fig:fig2}
\end{figure}

\begin{figure}
\centering
\includegraphics[scale=0.37]{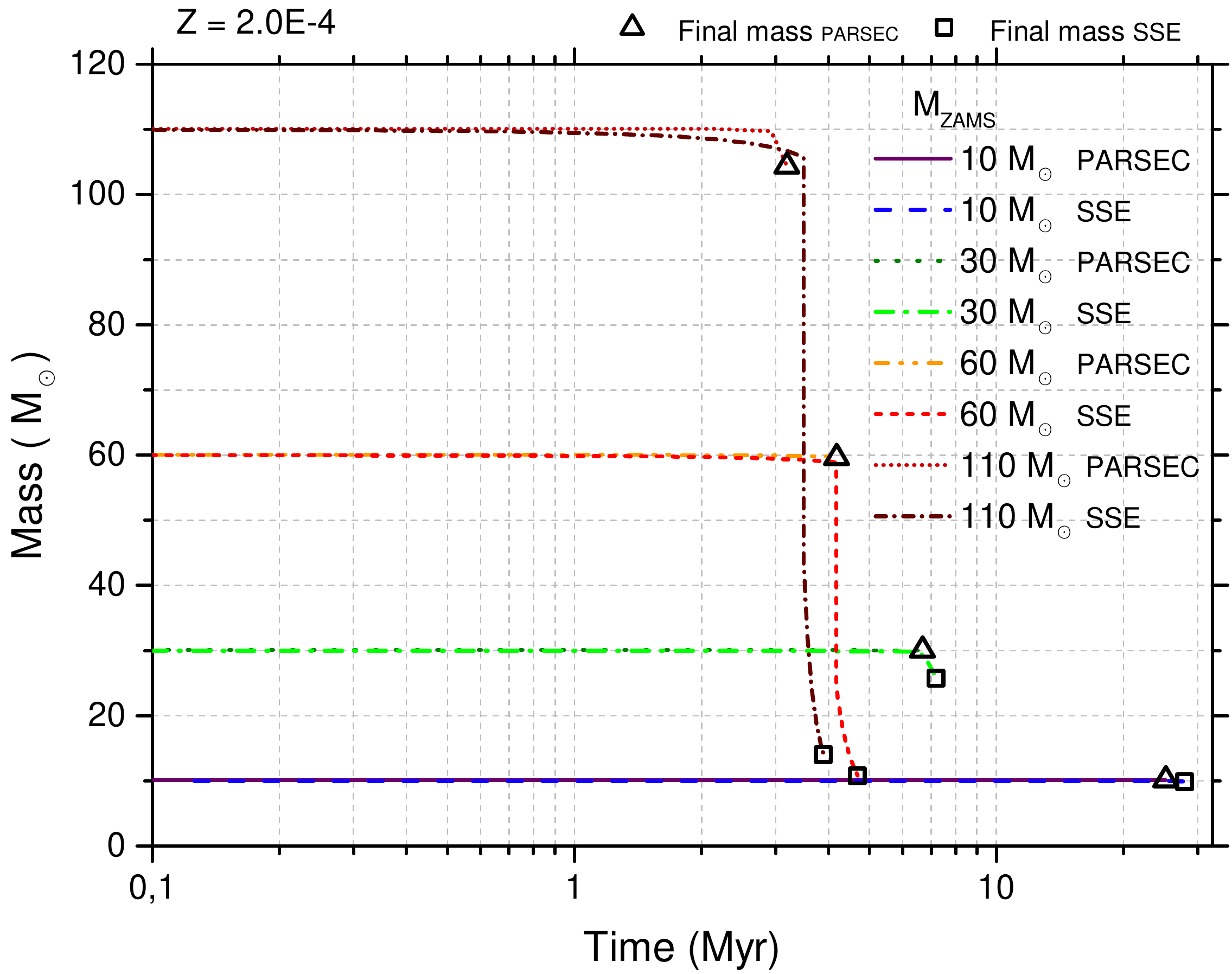}
\caption{Same as Fig. \ref{fig:fig1}, but for $Z=2.0\times{10^{-4}}$.}
\label{fig:fig3}
\end{figure}

 Figures~\ref{fig:fig1}, \ref{fig:fig2} and \ref{fig:fig3} show the temporal evolution of 
stellar mass  at $Z=2\times{}10^{-2}$, $2\times{}10^{-3}$ and $2\times{}10^{-4}$, respectively, for four 
selected ZAMS masses between 10 and 110 M$_\odot{}$. The evolution of the stellar mass 
predicted by \parsec{} is compared with that implemented in \sse{}. At lower ZAMS masses ($M_{\rm 
ZAMS}\lesssim{}10$ M$_\odot$) the behaviour of \parsec{} and \sse{} is almost indistinguishable. 

  For larger masses, there is no significant difference for most of the star's life, but there is a 
  significant difference in the final masses \mfin{}, especially at low metallicity. The differences in 
  \mfin{} are about 80\% of $M_{\rm ZAMS}$ for stars with $M_{\rm ZAMS}\gtrsim{}60$ M$\odot$ at 
  $Z=2\times{}10^{-4}$. The reason of these differences is the treatment of stellar winds, especially in the 
  late-MS, LBV and WR stages (see Section~\ref{subsec:parsec_description} and \citealt{bressan2012,tang2014} 
  for details).

\subsection{Final mass (\mfin{})  and  CO core mass (\mco{})}
The SN explosion mechanisms discussed by \citet{fryer2012}, and implemented in \codename{}, depend on the final
mass of the star, \mfin, and on its final CO mass, \mco{}  ( see equations \ref{eq:startrack}, \ref{eq:rapid} 
and \ref{eq:delayed}). Since both \mfin{} and \mco{} depend on the
initial mass of the star, \mzams{}, and on its metallicity, $Z$, thus also \mrem{} will depend on \mzams{} and
$Z$.
This implies that the mass spectrum of compact remnants strongly depends on the prescriptions adopted to
evolve the star until its
pre-SN stage.

\begin{figure}
\centering
\includegraphics[scale=0.37]{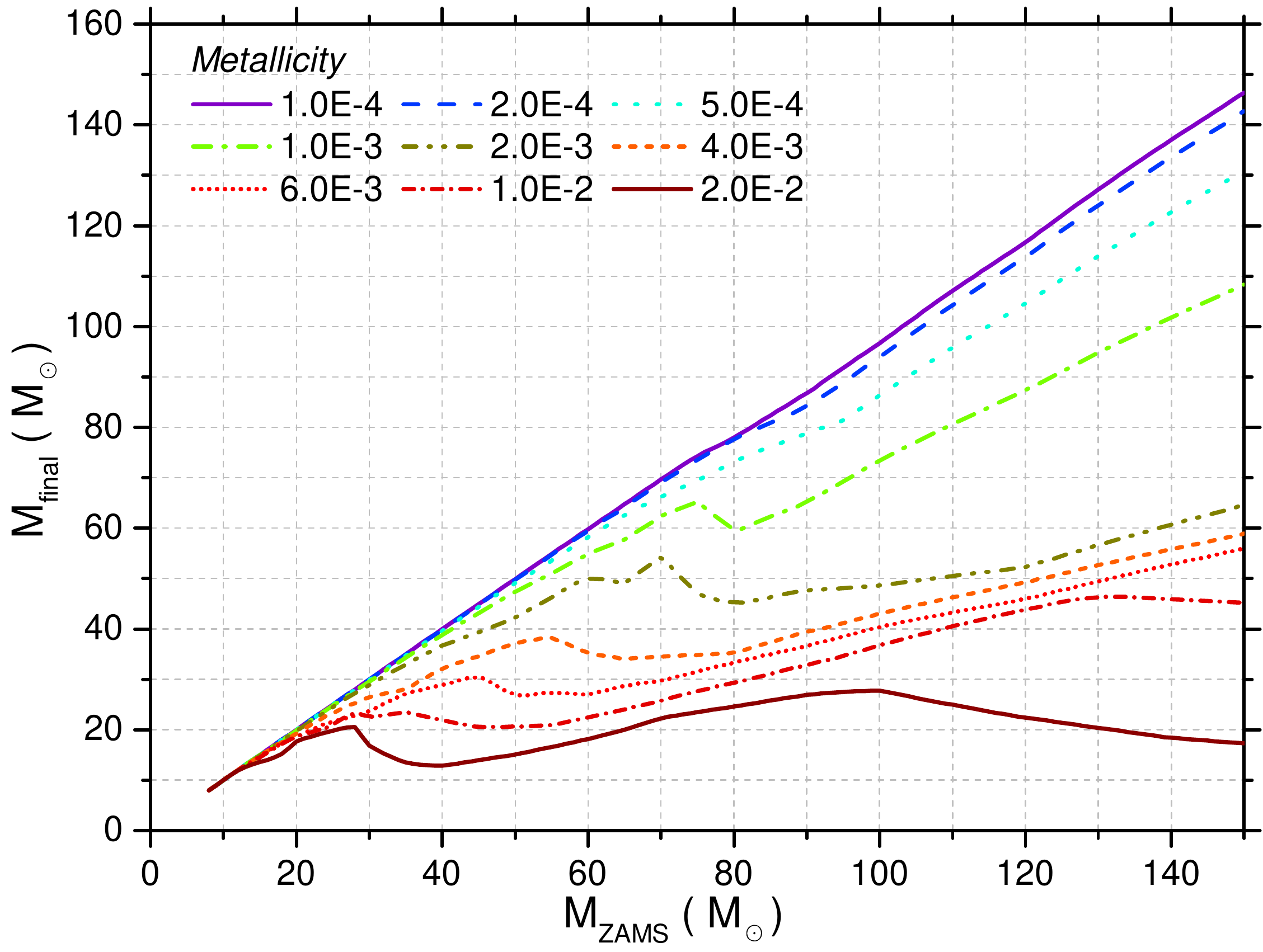}
\caption{Final mass of the stars as a function of their initial mass, for different values of
metallicity $1.0\times{10^{-4}}\leq Z \leq 2.0\times{10^{-2}}$. Top solid line: $Z=1.0\times{10^{-4}}$;
dashed line: $Z=2.0\times{10^{-4}}$; dotted line: $Z=5.0\times{10^{-4}}$; dash-dotted line:
$Z=1.0\times{10^{-3}}$; dash-double dotted line: $Z=2.0\times{10^{-3}}$; short dashed line:
$Z=4.0\times{10^{-3}}$; short dotted line: $Z=6.0\times{10^{-3}}$; short dash-dotted line:
$1.0\times{10^{-2}}$; bottom solid line: $Z=2.0\times{10^{-2}}$.}
\label{fig:fig4}
\end{figure}

Figure \ref{fig:fig4} shows the trend of \mfin{} as a function of  \mzams{}, for different values of the 
metallicity. Figure
\ref{fig:fig4} reflects the fact that metal-poor stars are subject to weaker stellar winds throughout 
their evolution. In fact, $\mfin$ is always smaller than $\sim 25\msun$ at
$Z=2.0\times{10^{-2}}$, while $\mfin\approx{}\mzams$ at $Z\lesssim{}2.0\times{10^{-4}}$. 
The curves for $Z\lesssim
2.0\times{10^{-4}}$ are well approximated by a simple linear relation

\begin{equation}
\mfin\left(\mzams\right) = 0.9519\mzams+1.45.
\label{eq:fitfinalmasses}
\end{equation}

\begin{figure}
\centering
\includegraphics[scale=0.37]{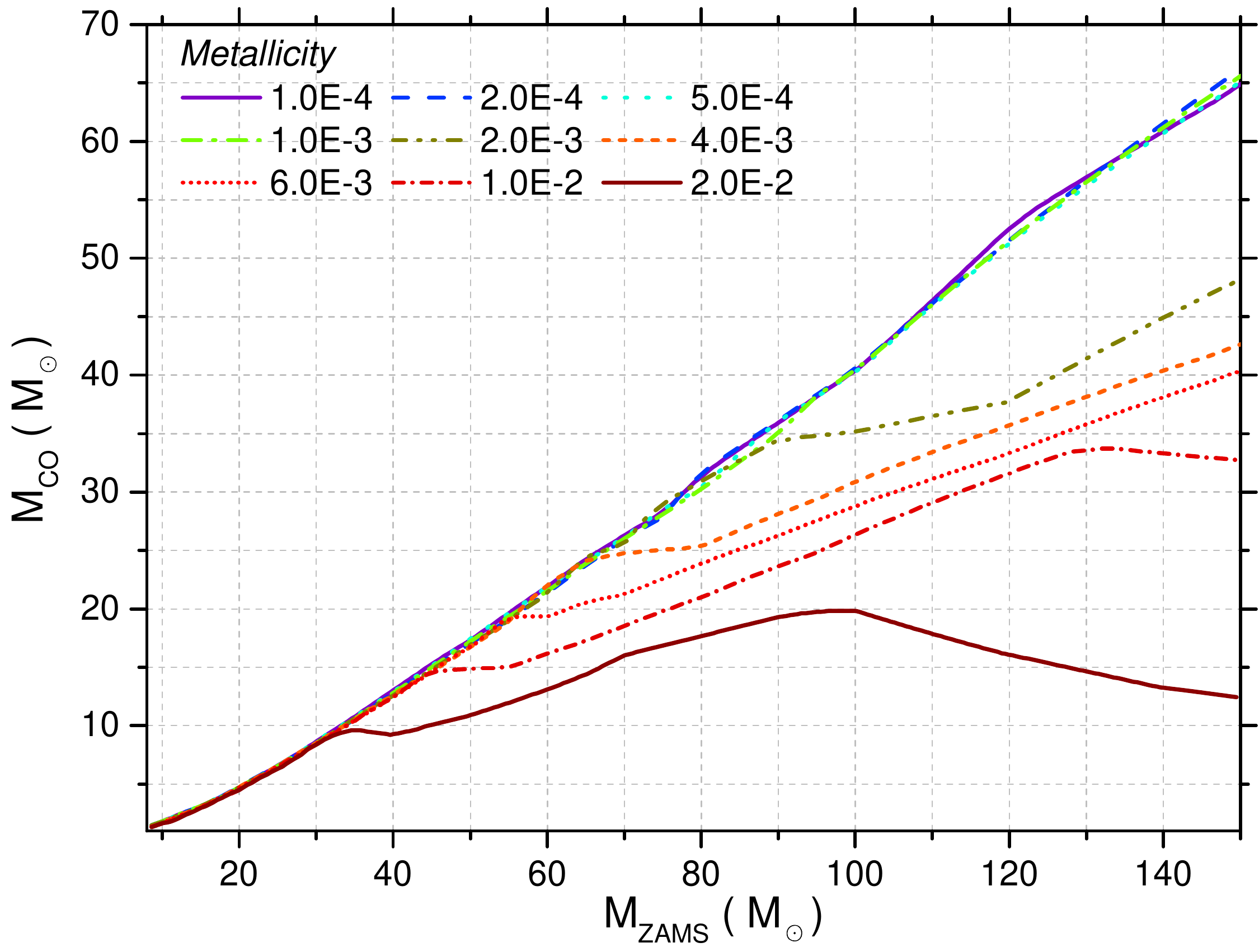}
\caption{Final mass of the CO core as a function of the initial mass of the star, for different values of
metallicity $1.0\times{10^{-4}}\leq Z \leq 2.0\times{10^{-2}}$. Line types are the same as in
Fig.~\ref{fig:fig4}.}
\label{fig:fig5}
\end{figure}

In Fig. \ref{fig:fig5}, we show \mco{} as a function of  \mzams,
for different values of metallicity. As expected, the final CO mass scales inversely with
metallicity: the maximum value of \mco{} ranges between $\sim20\msun$ and
$\sim65\msun$, for  $1.0\times{10^{-4}}\leq Z \leq 2.0\times{10^{-2}} $.
It is interesting to note that, for $Z\leq 1.0\times{10^{-3}}$, the curves of Fig. \ref{fig:fig5} become 
approximately independent of $Z$, and  can be expressed as
\begin{equation}
\mco\left(\mzams\right)=
\begin{cases}
0.3403\mzams-2.064\\
\text{\hspace{15pt}if } \mzams < 27\msun\\
0.4670\mzams-5.47\\
\text{\hspace{15pt}if } \mzams \geq 27\msun
\end{cases}
\label{eq:fitCO}
\end{equation}
At present, since \parsec{} does not include the TP-AGB stellar evolution phase,
equation~\ref{eq:fitCO} holds for
$\mzams \gtrsim M_{up} = 7 \msun$ (see Sec. \ref{subsec:stevoimpl}).

\subsection{The mass spectrum of compact remnants}
\begin{figure}
\centering
\includegraphics[scale=0.37]{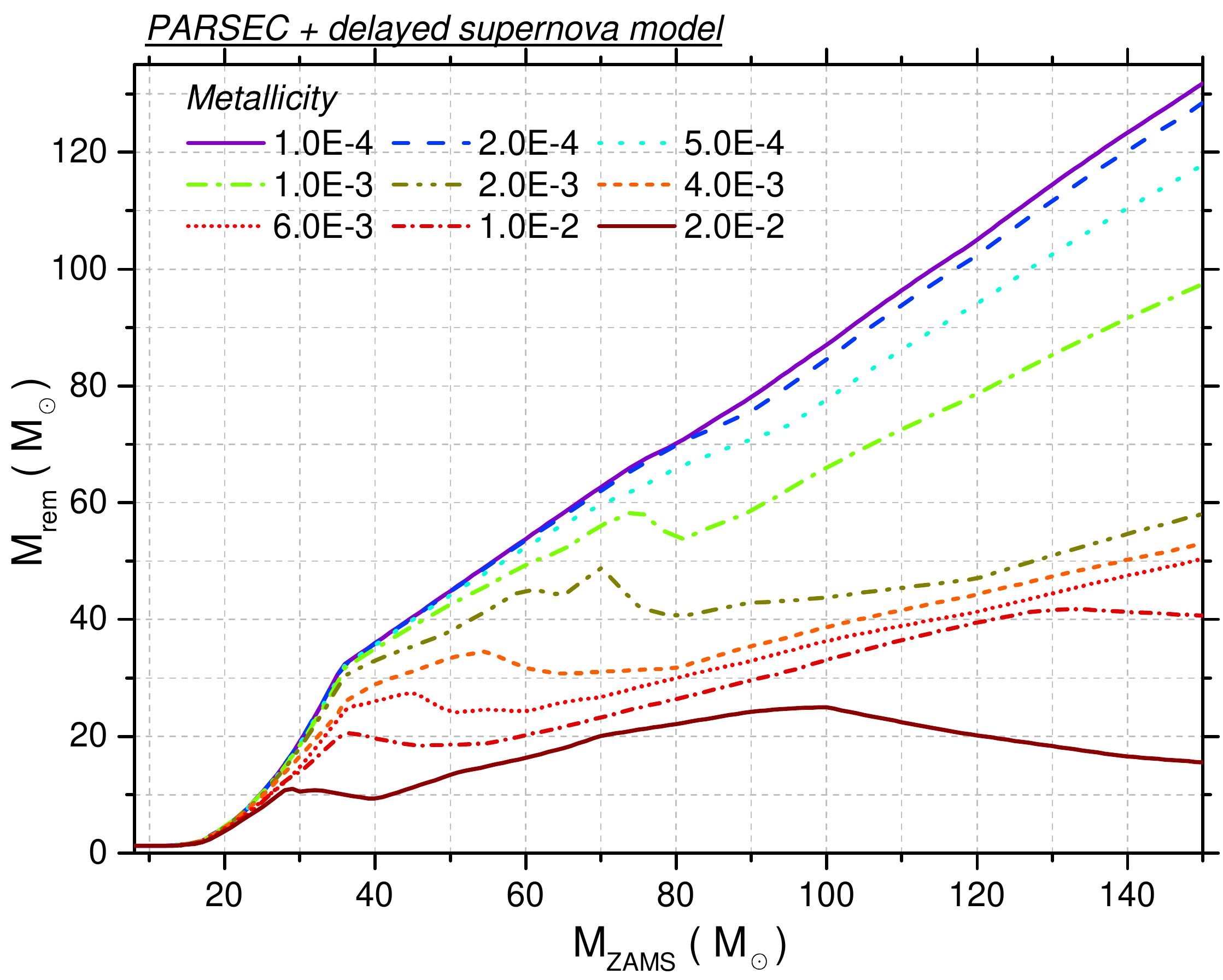}
\caption{Mass of the final compact remnant ($M_{\rm rem}$) as a function of the initial mass of the star, for various
metallicities. The curves have been obtained using \codename{} and the delayed SN model. Line types
are the same as in Fig. \ref{fig:fig4}.}
\label{fig:fig6}
\end{figure}
In Fig. \ref{fig:fig6}, we show the mass spectrum of compact remnants as a function of the ZAMS mass of their 
progenitors, for different values of metallicity. To obtain the curves in Fig. \ref{fig:fig6}, we used the 
delayed supernova model,  chosen as the default explosion
mechanism in \codename{}. As expected, in Fig. \ref{fig:fig6}, we notice that
the lower the metallicity is, the higher the mass of the heaviest compact remnant;
in particular, $\mrem$ ranges from $\sim 25 \msun$ at $Z=2.0\times{10^{-2}}$ to $\sim 135 \msun$ at
$Z=1.0\times{10^{-4}}$.
For $Z \lesssim 2.0\times{10^{-4}}$ and $7\msun =M_{{\rm up}}\leq \mzams \leq 150\msun$, simple fitting 
formulas can be derived for
$\mrem\left(\mzams\right)$, by substituting the best fit curves for
$\mfin\left(\mzams,Z\right)$ and $\mco\left(\mzams,Z\right)$ (Eqs. \ref{eq:fitfinalmasses} and
\ref{eq:fitCO}, respectively) in the formulas of the
delayed explosion mechanism (Eq. \ref{eq:delayed}):
\begin{equation}
M_{rem,bar}=
\begin{cases}
1.4 \msun \\
 \text{\hspace{15pt} if } M_{up}\lesssim \mzams \lesssim 13 \msun \\
 \\
0.170\mzams-0.882 \\
 \text{\hspace{15pt} if } 13\msun \lesssim \mzams \lesssim 16\msun \\
 \\
(0.041\mzams^3 - 0.673\mzams^2 + \\
+ 2.18\mzams + 0.361)/\left(0.952\mzams+0.15\right)  \\
  \text{\hspace{15pt} if } 16\msun\lesssim \mzams \lesssim 27\msun \\
  \\
(0.0563\mzams^3 - 1.10\mzams^2 + \\
+ 2.49\mzams + 0.318)/\left(0.952\mzams+0.15\right)  \\
  \text{\hspace{15pt} if } 27\msun\lesssim \mzams \lesssim 36\msun \\
  \\
0.952\mzams + 1.45 \\
 \text{\hspace{15pt} if } \mzams \geq 36\msun.
\end{cases}
\label{eq:fitremnant}
\end{equation}
A general fitting formula for \mrem{}, as a function of \mzams{} and $Z$, and that holds for every 
metallicity, is provided in Appendix~\ref{appsec:fittingformula}.

Figure \ref{fig:fig7} shows the value of \mrem{} as a function of \mco{}, for different metallicities.
It is worth noting that, for every metallicity, \mrem{} lies approximately between $\mrem{}_{\rm ,up}=
1.85\mco + 11.9$ and
$\mrem{}_{\rm ,down}= 1.22\mco+1.06$ (see Appendix \ref{appsec:fittingformula} for the details).

\begin{figure}
\centering
\includegraphics[scale=0.37]{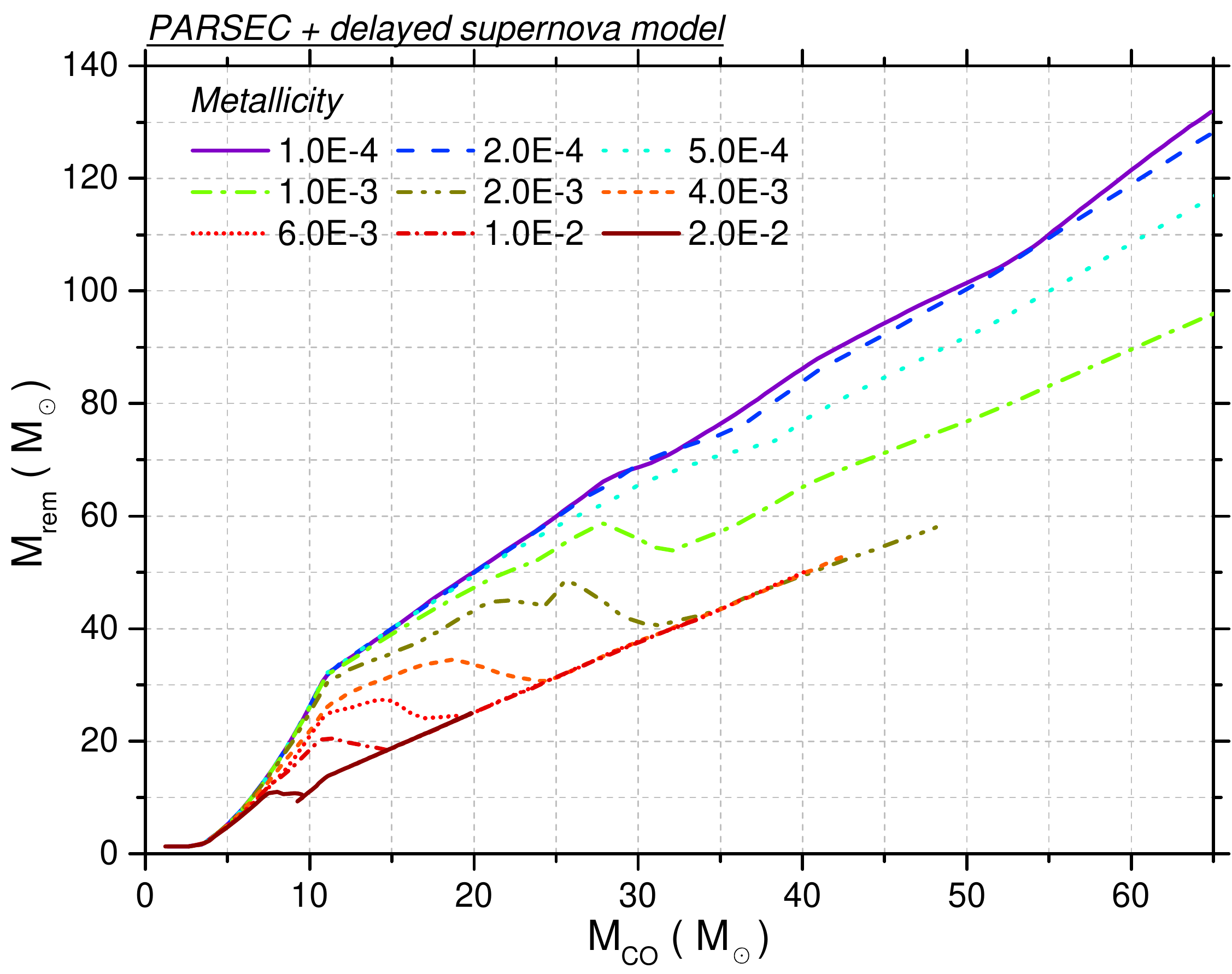}
\caption{Mass of the compact remnant as a function of the final CO core mass of the progenitor, for
different metallicities. The curves have been obtained using \codename{} and the delayed SN model. Line types 
are the same as in Fig. \ref{fig:fig4}.}
\label{fig:fig7}
\end{figure}

\subsection{Comparison of different supernova explosion models}

\begin{figure}
\centering
\includegraphics[scale=0.37]{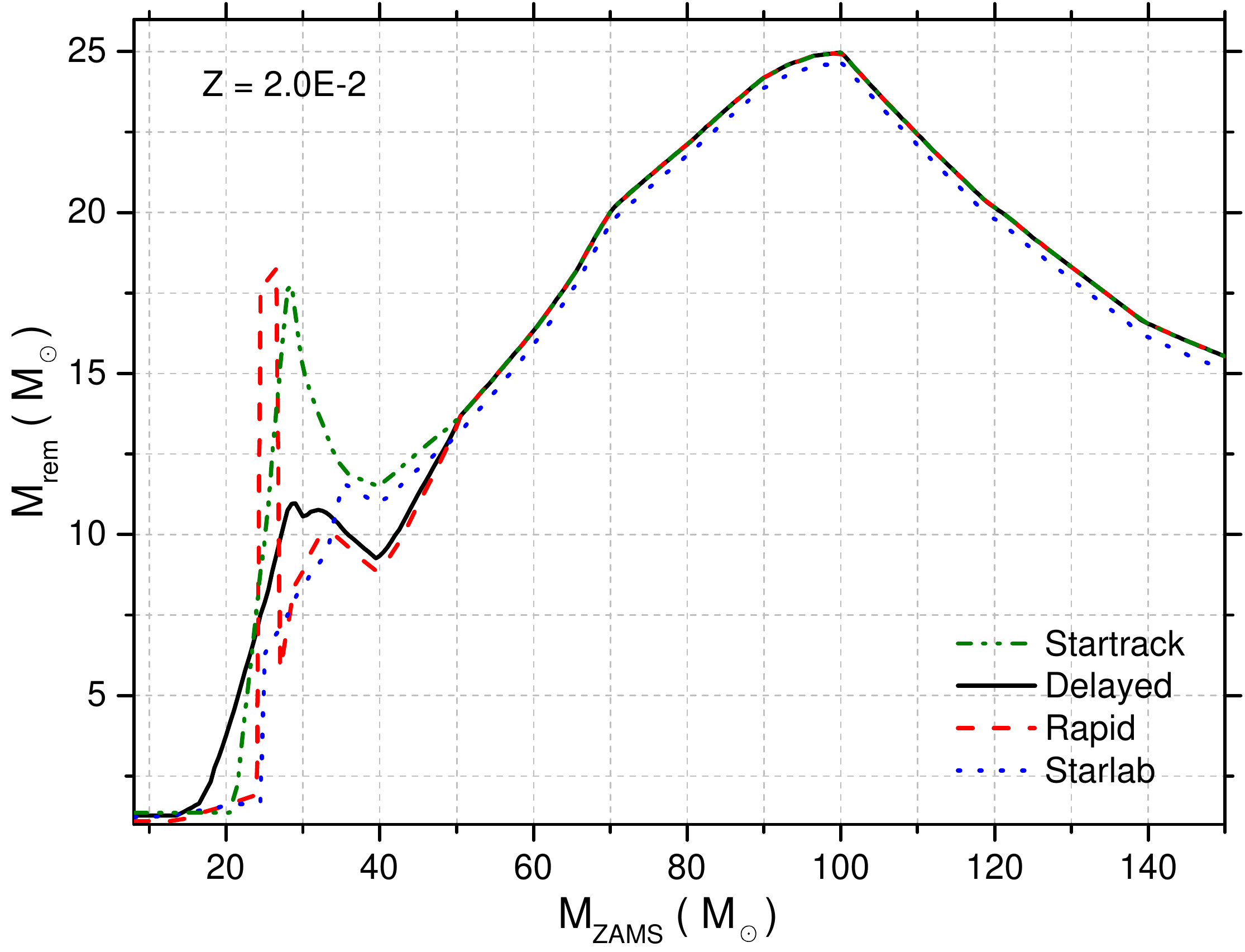}
\caption{Mass of compact remnants as a function of $M_{\rm ZAMS}$ at $Z=2.0\times{10^{-2}}$, derived with 
\codename{} using different models of SN explosion. Dash-double
dotted line: \startrack{} SN recipes; solid line: delayed SN model; dashed line: rapid SN model; dotted line: 
\starlab{} prescriptions.}
\label{fig:fig8}
\end{figure}

In  Fig. \ref{fig:fig8},  we show the mass of the remnants as a function of
\mzams, for different SN recipes, at fixed metallicity $Z=2.0\times{10^{-2}}$, in order to compare the various SN models
implemented in \codename{}. In this figure, we also show the results obtained using the \seba{} (\starlab{}) 
built-in models(see \citealt{zwart2001} for details).

\begin{figure}
\centering
\includegraphics[scale=0.37]{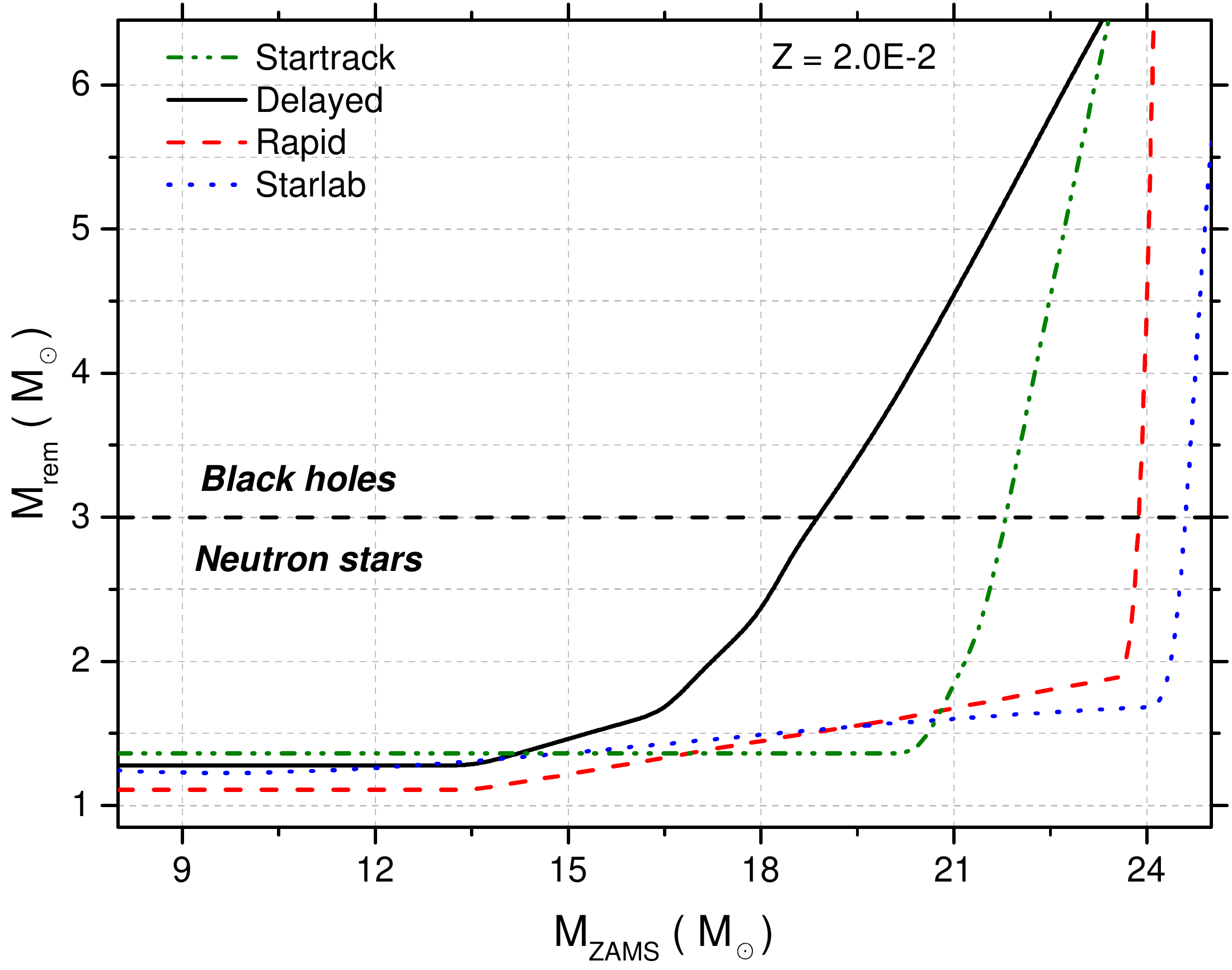}
\includegraphics[scale=0.37]{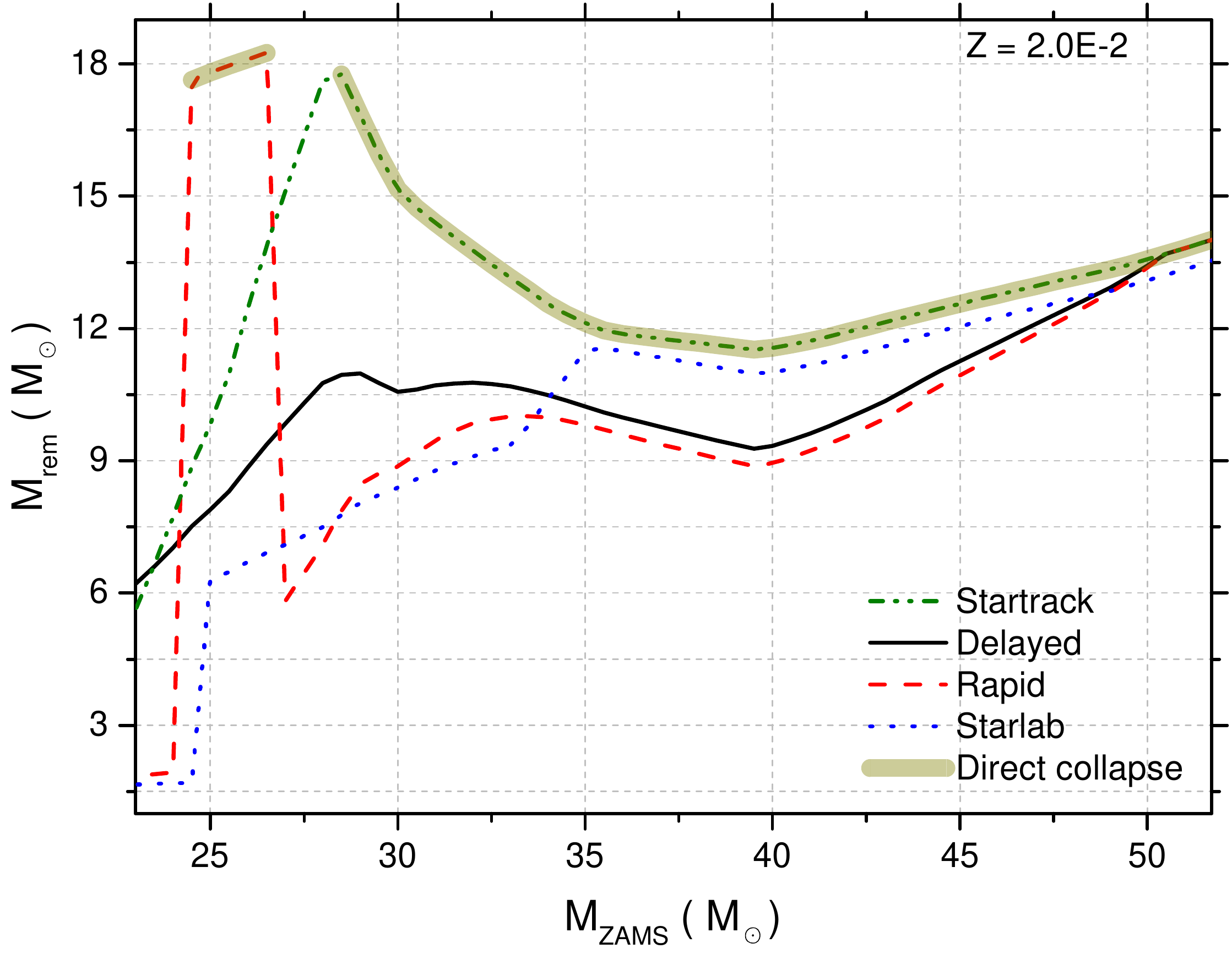}
\caption{Two details of Fig. \ref{fig:fig8}: the top panel shows the range $8\msun \lesssim \mzams
\lesssim 25 \msun$, while the bottom panel refers to the interval $25\msun \lesssim \mzams
\lesssim 50 \msun$. In the top panel, the horizontal dashed line marks the transition between NSs and BHs.
The thick, semi-transparent line in the bottom panel highlights the intervals in which direct collapse 
occurs. The other lines are the same as in Fig. \ref{fig:fig8}.}
\label{fig:fig9}
\end{figure}

Figure \ref{fig:fig8} shows that all recipes produce approximately the
same remnant mass spectrum, for $\mzams\gtrsim 50\msun$. The bottom panel of Fig. \ref{fig:fig9} is
a zoom of Fig. \ref{fig:fig8} in the region of $25\msun \le{}\mzams\le{}50\msun$. From Fig. \ref{fig:fig9} we 
notice that the \startrack{} SN recipes produce, on average, more massive BHs (with mass between $\sim{} 
12\msun$
and $\sim{} 18\msun$) in the interval $28\msun \lesssim \mzams \lesssim 50 \msun$. This is due to the fact
that
\startrack{} predicts the formation of compact remnants via direct collapse if $\mco\geq 7.6\msun$
(see equation \ref{eq:startrack}), condition that occurs for
$\mzams\gtrsim 28\msun$ (see Fig. \ref{fig:fig5}). The other models do not predict direct collapse in
this interval of $\mzams$, and produce lighter BHs with masses
between $\sim 6 \msun$ and $\sim 13 \msun$.

The abrupt step of the rapid SN model, for $24\msun\lesssim\mzams\lesssim26\msun$, corresponds to the
process of direct collapse that takes place for $6\msun\leq\mco\leq7\msun$ in this model (see equation~\ref{eq:rapid}).

In the range $14\msun \le{} \mzams \le{} 24 \msun$ (see upper panel of Fig. \ref{fig:fig9}), the delayed 
model predicts a higher amount of
fallback than the other models. In fact, the delayed mechanism forms compact objects with masses between
$\sim 2.0 \msun$ and $\sim 6.0 \msun$, while the other models form remnants with masses only up to $\sim 2
\msun$ (see upper panel of Fig. \ref{fig:fig9}). Using the \startrack{} prescriptions, it is possible
to form remnants with masses $\gtrsim 3.0 \msun$, but only for $\mzams \gtrsim 22\msun$. Finally, using
the SN model implemented in \seba{} and the rapid SN model, we find a paucity of remnants with masses between $\sim
2\msun$ and $\sim 6\msun$ with the result of having a marked gap between the heaviest NS and
the lightest BH.

\begin{figure}
\centering
\includegraphics[scale=0.37]{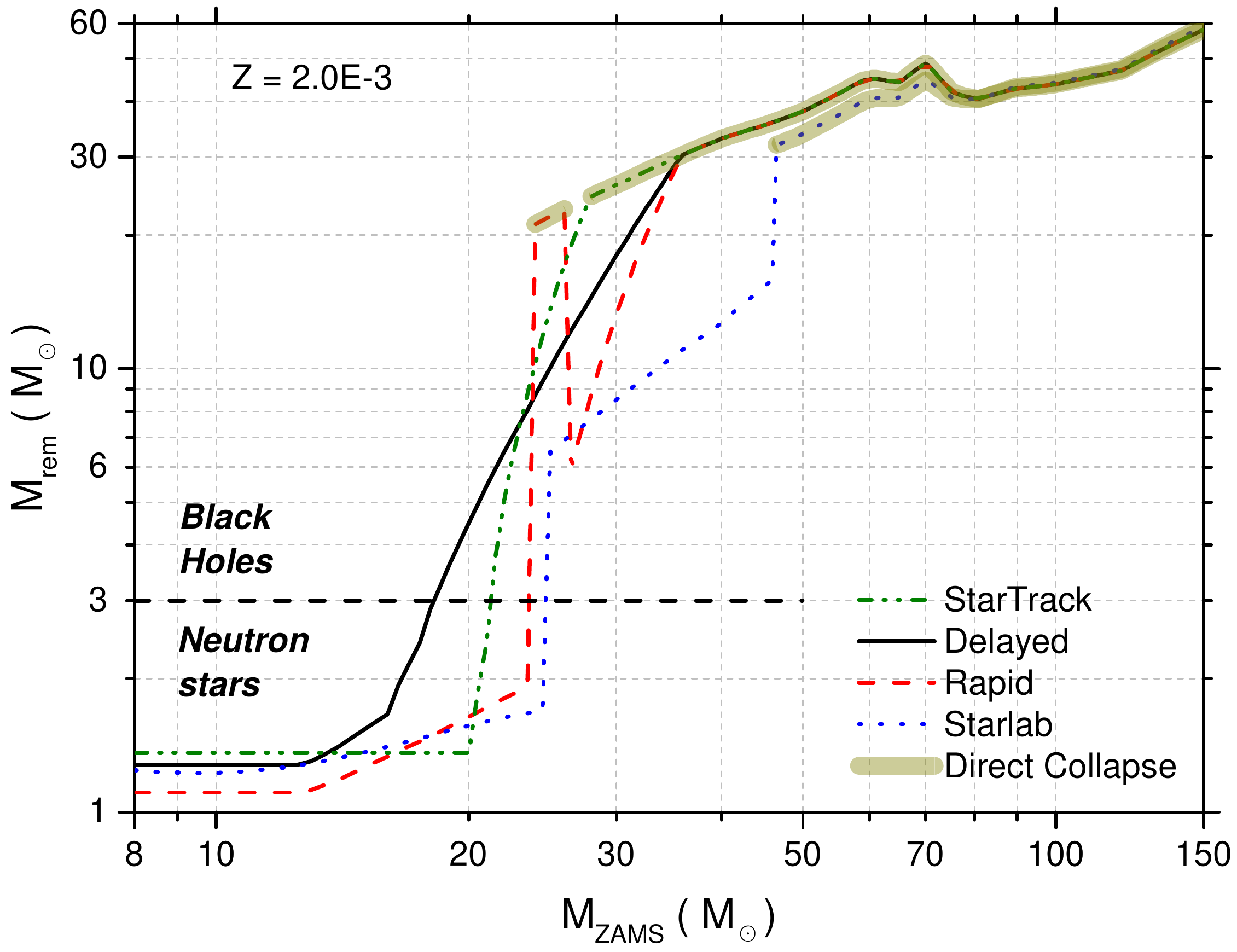}
\caption{Mass of compact remnants as a function of $M_{\rm ZAMS}$ at $Z=2.0\times{10^{-3}}$, derived with 
\codename{} using different models of SN explosion. Line types are the same as in Fig. \ref{fig:fig9}.}
\label{fig:fig10}
\end{figure}

\begin{figure}
\centering
\includegraphics[scale=0.37]{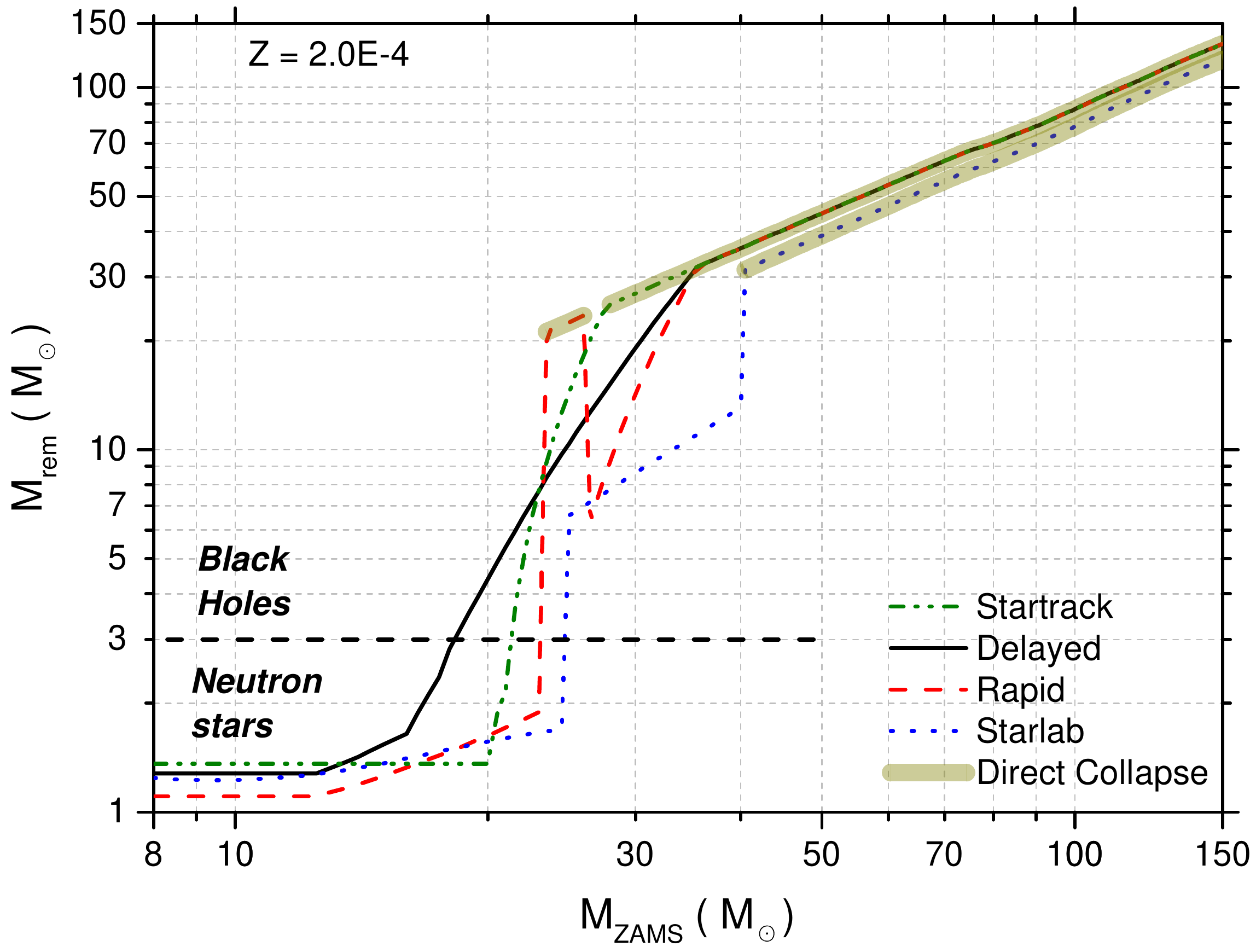}
\caption{ The same as Fig. \ref{fig:fig10}, but for $Z=2.0\times{10^{-4}}$.}
\label{fig:fig11}
\end{figure}

In Figs. \ref{fig:fig10} and \ref{fig:fig11}, we show the
mass spectrum of BHs and NSs obtained for different explosion models for
$Z=2.0\times{10^{-3}}$ and $Z = 2.0\times{10^{-4}}$, respectively.
At $Z=2.0\times{10^{-3}}$ ($Z = 2.0\times{10^{-4}}$), the maximum BH mass is $\sim 60\msun$ ($\sim 130 
\msun$), regardless of the SN
explosion mechanism. The main remarkable features of  Figures \ref{fig:fig10} and 
\ref{fig:fig11} are the following:
\begin{enumerate}
\item the \startrack{} models produce heavier compact remnants for $25\msun\le{}\mzams\le{}35\msun$;
\item the rapid SN model exhibits an abrupt step for $24\msun\le{}\mzams\le{}26\msun$;
\item for $\mzams\gtrsim 35\msun$, the mass spectra obtained with the models of \citet{fryer2012} become
indistinguishable (all of them predict direct collapse);
\item except for the delayed model, we obtain a paucity of remnants with masses between $\sim 2\msun$ and
$\sim 6\msun$;
\item The \seba{} built-in SN explosion model predicts direct collapse for $\mzams \gtrsim 45\msun$ ($\mzams 
\gtrsim 40\msun$) at $Z =
2.0\times{10^{-3}}$ ($Z = 2.0\times{10^{-4}}$).
\end{enumerate}

In Section~\ref{sec:newmodels}, we extend this comparison to more sophisticated models of SN explosion 
(based on the compactness of the stellar core at pre-SN stage, and on the dimensionless entropy per nucleon 
at pre-SN stage).

\subsection{Comparisons with other stellar evolution tools}
\begin{figure}
\centering
\includegraphics[scale=0.37]{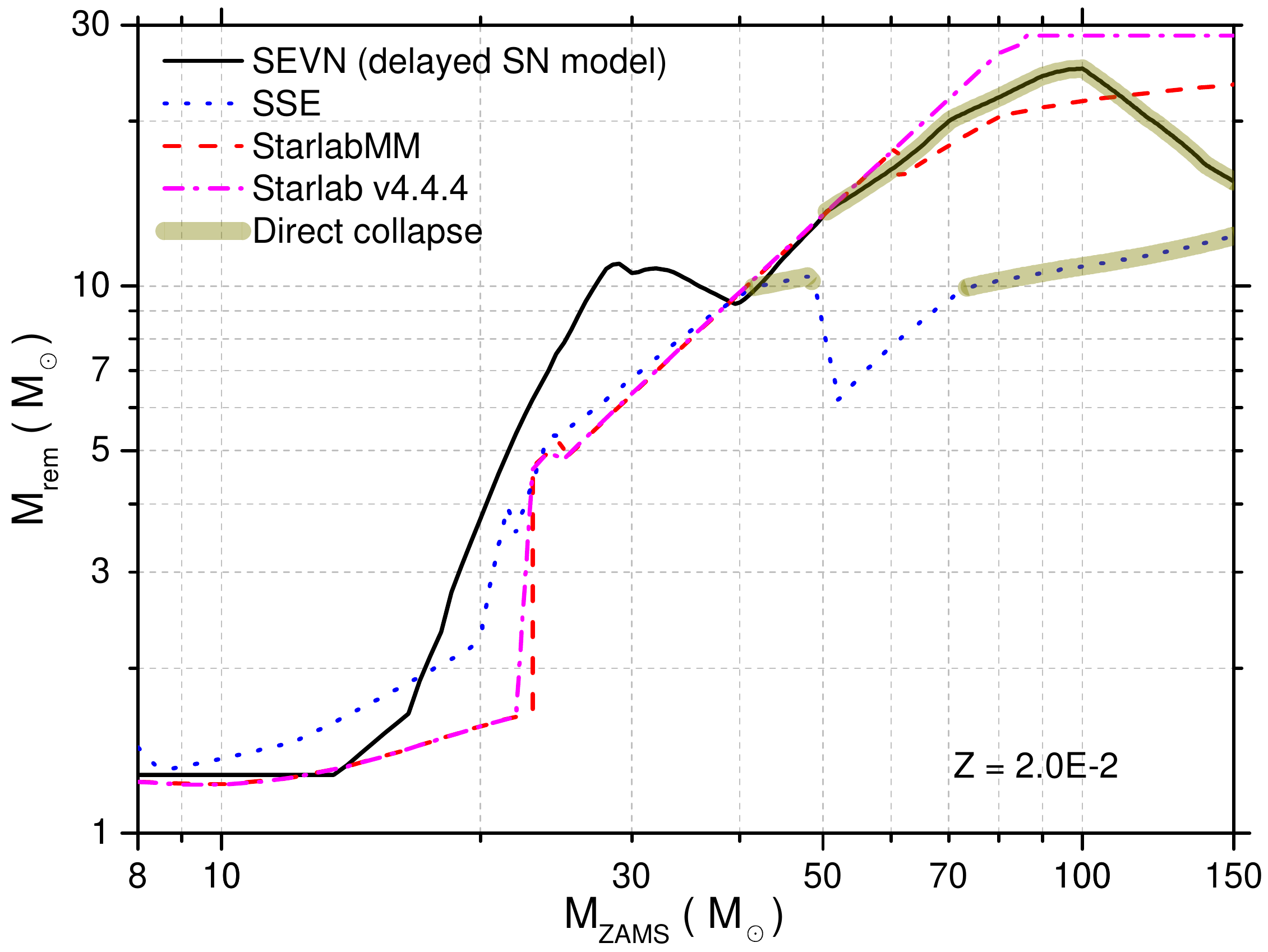}
\caption{ Mass of compact remnants as a function of $M_{\rm ZAMS}$ at $Z=2.0\times{10^{-2}}$, derived with 
different codes: \codename{} (solid line), \sse{} (dotted line), \starlabmm{} (dashed line), \starlab{} 
v4.4.4 (dash-dotted line). The
semi-transparent line highlights the intervals in which direct collapse takes place. For \codename{}, we used the delayed SN mechanism. 
\starlabmm{} is the modified version of \starlab{} described in \citet{mapelli2013}, while \starlab{} v4.4.4 
is the standard version of \starlab{} (ver. 4.4.4).}
\label{fig:fig12}
\end{figure}

Figure \ref{fig:fig12} shows the mass spectrum of compact remnants, at $Z=2.0\times{10^{-2}}$,
obtained using \codename{}, with the delayed supernova explosion model, in comparison with the results
of
 \starlab{} v4.4.4 (default \seba{} stellar evolution module, \citealt{zwart2001}, hereafter simply \starlab{}), \starlabmm{}
(\citealt{mapelli2013}) and \sse{}. The maximum BH
mass we obtain, at
$Z=2.0\times{10^{-2}}$, using \codename{}, is
$\sim 25\msun$, while using \starlab{} this value is slightly higher ($\sim 28 \msun$). In \codename{}, the
stars with $\mzams\simeq 100 \msun$ form the heaviest BHs, while, using \starlab{}, the most massive
remnants derive from stars with $85\msun \le{} \mzams \le{} 150 \msun$. \starlabmm{} produces BHs with masses up to $\sim 23 \msun$. It is interesting to
point out that the recipes implemented in \starlab{} produce a paucity of compact remnants with masses
between $\sim 2\msun$ and $\sim 5 \msun$. This gap derives from the assumption that BHs form only if $\mco\geq 5 \msun$, otherwise, NSs with masses between $\sim 1.2\msun$ and $\sim 1.6 \msun$ are formed.
If we use the
\sse{} package, the maximum mass of compact remnants is $\sim 13 \msun$. It is also important
to stress that, for $17 \msun \lesssim \mzams
\lesssim 40 \msun$, the delayed explosion model implemented in \codename{}
creates more massive compact remnants than the other models.

\begin{figure}
\centering
\includegraphics[scale=0.37]{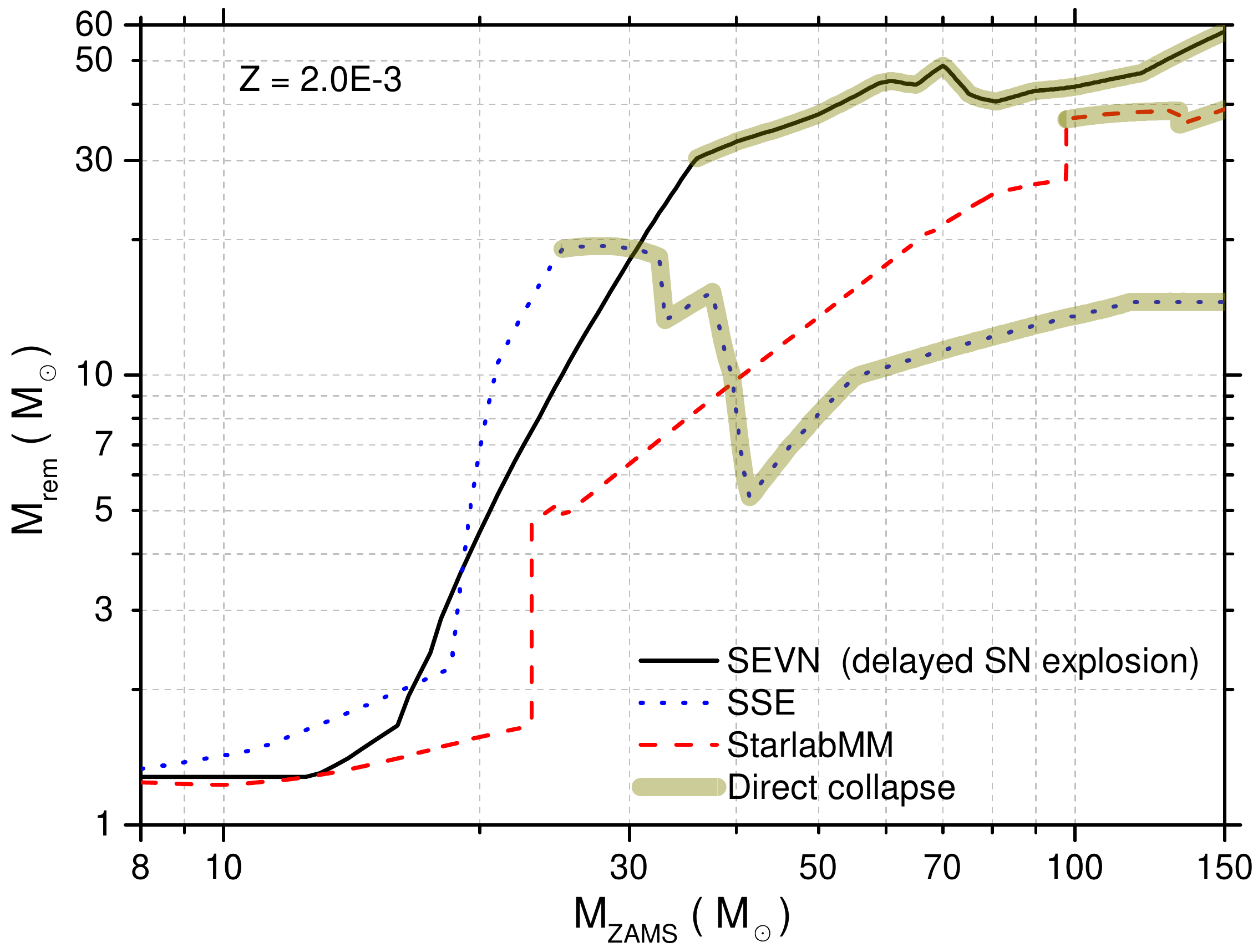}
\caption{ Mass of compact remnants as a function of $M_{\rm ZAMS}$ at $Z=2.0\times{10^{-3}}$, derived with 
different codes: \codename{} (solid line), \sse{} (dotted line), \starlabmm{} (dashed line). The
semi-transparent line highlights the intervals in which direct collapse takes place. For \codename{}, we used 
the delayed SN mechanism. }
\label{fig:fig13}
\end{figure}

\begin{figure}
\centering
\includegraphics[scale=0.37]{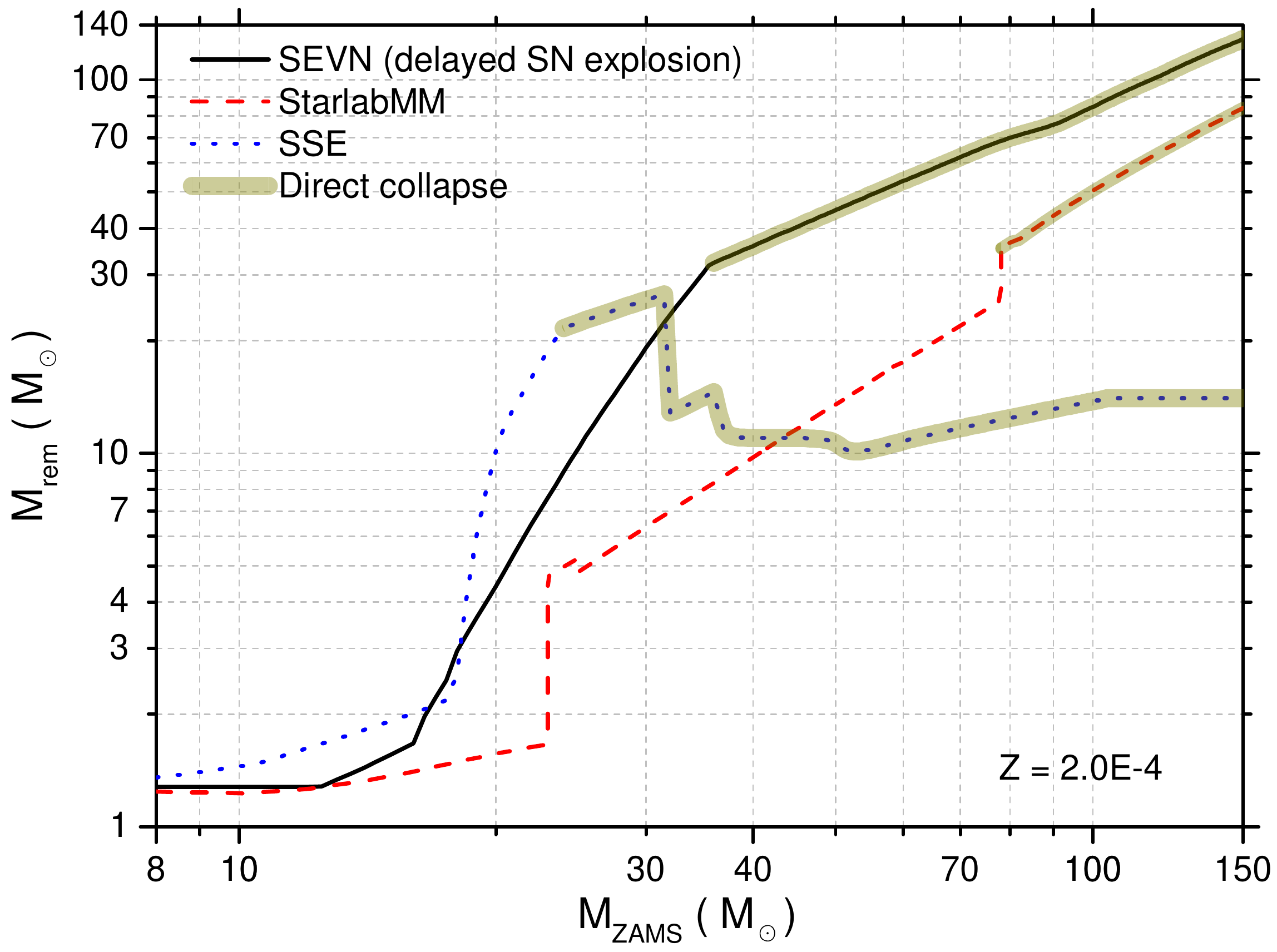}
\caption{ The same as Fig.~\ref{fig:fig13} but for $Z=2.0\times{10^{-4}}$.}

\label{fig:fig14}
\end{figure}

Figures \ref{fig:fig13} and  \ref{fig:fig14} show the mass spectrum of compact
remnants at $Z=2.0\times{10^{-3}}$ and $Z=2.0\times{10^{-4}}$, respectively.
The results of \starlab{} are not shown in Figures \ref{fig:fig13} and  \ref{fig:fig14}, because 
\starlab{} does not include metallicity dependent stellar winds. At $Z=2.0\times{10^{-3}}$ and for $\mzams 
\gtrsim 30\msun$, \codename{} (with the \parsec{} evolutionary tables) produces significantly heavier BHs 
than \starlabmm{} and \sse{} (see Fig. \ref{fig:fig13}). In particular, the maximum BH
mass obtained using \codename{} is $\sim 60\msun$, while this value is  $\sim 40 \msun$  and $\sim 20\msun$ in the
case of \starlabmm{} and \sse{}, respectively. We also stress that,
while
for \codename{} and \starlabmm{} the heaviest BH comes from the death of the most massive star (that is
$\mzams = 150 \msun$), in the case of \sse{} BHs of $\sim 20 \msun$ form from stars with
$25\msun \lesssim \mzams \lesssim 30\msun$ only. The abrupt step observed for $\mzams\simeq 100\msun$, in the \starlabmm{} curve represents the transition between partial fallback and direct collapse (occurring at $\mfin \geq 40
\msun$, see \citealt{mapelli2013} for details), while that at $\mzams \simeq 25 \msun$ reflects the
transition from
NSs to BHs. It is worth noting that, for $Z=2.0\times{10^{-3}}$ and $\mzams \lesssim 30
\msun$, the \sse{} model produces
more massive compact remnants than the other models.

Similar considerations hold for Fig. \ref{fig:fig14}, which is the same as Figs. \ref{fig:fig12}
and \ref{fig:fig13} but at $Z=2.0\times{10^{-4}}$. \codename{} predicts BHs masses up to $\sim
120 \msun$, \starlabmm{} creates BHs of maximum mass $\sim 80 \msun$, while the
\sse{} prescriptions do not go beyond $\sim 25 \msun$. Also in this case, as observed at
$Z=2.0\times{10^{-3}}$, the \sse{} recipes predict the formation of more massive compact remnants in the
range $20\msun\lesssim\mzams \lesssim 30 \msun$.

\begin{table}
\begin{center}
\caption{Values of \mzams{}, \mfin{} and \mco{} that correspond to the transition 
between the formation of a NS and that of a BH, and  maximum BH mass ($M_{\mathrm{BH}}^{\mathrm{max}}$), for 
three different codes: \codename{}, \starlabmm{} and \sse{}. D: delayed model; R: rapid model; S: 
\startrack{} prescriptions. Results for $Z=2.0\times 10^{-2}$.}
\label{tab:tab1}
\begin{tabular*}{\columnwidth}{@{\extracolsep{\fill}} cccccc}
\toprule
  &  \multicolumn{3}{c}{\codename{}} & \starlabmm{}     & \sse{}    \\
\cmidrule(r){2-4}
& D & R & S & &\\
\midrule
$\mzams$ & $18.8$ & $23.9$ & $21.8$ & $23.0$ & $20.7$ \\
$\mfin$ & $16.0$ & $19.6$ & $18.7$ & $17.3$ & $7.3$ \\
$\mco$ & $4.1$ & $6.0$ & $5.1$ & $5.0$ & $5.3$ \\\hline
$M_{\mathrm{BH}}^{\mathrm{max}}$ & $25.0$ & $25.0$ & $25.0$ & $23.0$ & $12.0$ \\
\bottomrule
\end{tabular*}
\end{center}
\end{table}

\begin{table}
\begin{center}
\caption{Same as Tab.\ref{tab:tab1} but for $Z=2.0\times 10^{-3}$.}
\label{tab:tab2}
\begin{tabular*}{\columnwidth}{@{\extracolsep{\fill}} cccccc}
\toprule
  &  \multicolumn{3}{c}{\codename{}} & \starlabmm{}     & \sse{}    \\
\cmidrule(r){2-4}
& D & R & S & &\\
\midrule
$\mzams$ & $18.2$ & $23.6$ & $21.3$ & $23.0$ & $18.8$ \\
$\mfin$ & $17.9$ & $23.2$ & $21.1$ & $17.6$ & $16.4$ \\
$\mco$ & $4.1$ & $6.0$ & $5.1$ & $5.0$ & $5.1$ \\ \hline
$M_{\mathrm{BH}}^{\mathrm{max}}$ & $58.0$ & $58.0$ & $58.0$ & $39.0$ & $19.0$ \\
\bottomrule
\end{tabular*}
\end{center}
\end{table}

\begin{table}
\begin{center}
\caption{Same as Tab.\ref{tab:tab1} but for $Z=2.0\times 10^{-4}$.}
\label{tab:tab3}
\begin{tabular*}{\columnwidth}{@{\extracolsep{\fill}} cccccc}
\toprule
  &  \multicolumn{3}{c}{\codename{}} & \starlabmm{}     & \sse{}    \\
\cmidrule(r){2-4}
& D & R & S & &\\
\midrule
$\mzams$ & $18.2$ & $23.1$ & $21.3$ & $23.0$ & $18.0$ \\
$\mfin$ & $18.0$ & $23.1$ & $21.3$ & $17.7$ & $17.0$ \\
$\mco$ & $4.1$ & $6.0$ & $5.1$ & $5.0$ & $5.1$ \\\hline
$M_{\mathrm{BH}}^{\mathrm{max}}$ & $130.0$ & $130.0$ & $130.0$ & $83.0$ & $26.0$ \\
\bottomrule
\end{tabular*}
\end{center}
\end{table}

Tables \ref{tab:tab1}, \ref{tab:tab2} and \ref{tab:tab3} report the values of 
\mzams{}, \mfin{} and \mco{}, corresponding to the transition between the formation 
of a NS and a BH, 
at $Z=2\times{}10^{-2}$, $2\times{}10^{-3}$ and 
$2\times{}10^{-4}$, respectively. The results obtained using \codename{}, 
\starlabmm{} and \sse{} are compared in the Tables. We notice that the transition value of \mco{} does not depend on metallicity and it 
ranges from $\sim 4.0\msun$ (delayed model of \codename{}) to $\sim 6.0\msun$ (rapid model of \codename{}). 
The transition values of \mzams{} and \mfin{} show a weak dependence on metallicity for a given code. 
\mzams{} goes form $\sim 18 \msun$ (delayed model of \codename{} at low metallicity) to $\sim 24 \msun$ 
(rapid model of \codename{} at $Z=2\times{}10^{-2}$), while \mfin{} ranges form $\sim 7\msun$ (\sse{} at 
$Z=2\times{}10^{-2}$) to $23\msun$ (rapid model of \codename{} at low metallicity). In the last row of tables 
\ref{tab:tab1}, \ref{tab:tab2} and \ref{tab:tab3}, we also report the maximum compact remnant 
mass. As we have already shown in this section, for the maximum BH mass we get huge differences between the considered
codes. This is due to the different stellar evolution recipes adopted in \parsec{}, \sse{} and \starlabmm{}, 
especially for metal-poor stars.

\subsection{The mass distribution of compact remnants}
In this section, we derive the mass function of compact remnants (NSs and BHs) that form in a stellar 
population following the Kroupa initial mass function (IMF, \citealt{kroupa2001}). The Kroupa IMF scales as 
${\rm d}N/{\rm d}m\propto{}m^{-\alpha{}}$,  with  $\alpha=1.3\,{}(2.3)$ for $m<0.5$ M$_\odot$ ($>0.5$ 
M$_\odot$). We assume a   minimum mass $m_{\mathrm{min}}=0.1\msun$ and a maximum mass 
$m_{\mathrm{max}}=150\msun$. We consider three different metallicities ($Z=2\times{}10^{-2}$, 
$2\times{}10^{-3}$ and $2\times{}10^{-4}$). For each metallicity, we generate $2.5\times{}10^6$ MS stars, 
with mass distributed according to the Kroupa IMF, and we evolve them with \codename{}. For each case, we do 
three realizations: one with the delayed SN model, one with the rapid SN model, and one with the Startrack 
recipes for compact remnants. Moreover, we also compare 
\codename{} (with the delayed SN recipe) with \starlabmm{} and with \sse{}. Table \ref{tab:tab4} lists the 
properties of the different realizations.

\begin{table}
  \begin{center}
  \caption{General properties of the stellar populations used as test-case to study the mass distribution of NSs 
    and BHs.}
  \label{tab:tab4}
  \begin{tabular}{llll}\hline 
 Run    & $Z$              & SN recipe & Code\\\hline 
 Z1D & $2\times{}10^{-2}$  & delayed  SN   &\codename{}\\
 Z2D & $2\times{}10^{-3}$  & delayed  SN     &\codename{} \\ 
 Z3D & $2\times{}10^{-4}$  & delayed  SN     &\codename{} \\
 Z1R & $2\times{}10^{-2}$  & rapid   SN    &\codename{} \\
 Z2R & $2\times{}10^{-3}$  & rapid   SN    &\codename{} \\ 
 Z3R & $2\times{}10^{-4}$  & rapid    SN   &\codename{} \\
 Z1S & $2\times{}10^{-2}$  & Startrack   SN    &\codename{} \\
 Z2S & $2\times{}10^{-3}$  & Startrack   SN      &\codename{} \\ 
 Z3S & $2\times{}10^{-4}$  & Startrack   SN      &\codename{} \\
 Z1MM &  $2\times{}10^{-2}$  & Mapelli et al. (2013) & \starlabmm{}\\
 Z2MM &  $2\times{}10^{-3}$  & Mapelli et al. (2013) & \starlabmm{}\\
 Z3MM &  $2\times{}10^{-4}$  & Mapelli et al. (2013) & \starlabmm{}\\
 Z1SSE & $2\times{}10^{-2}$  & Hurley et al. (2000) & \sse{}\\
 Z2SSE & $2\times{}10^{-3}$  & Hurley et al. (2000) & \sse{}\\
 Z3SSE & $2\times{}10^{-4}$  & Hurley et al. (2000) & \sse{}\\\hline 
  \end{tabular}
\footnotesize{We generated and evolved $2.5\times{}10^6$ stars in each of these runs.}
 \end{center}
\end{table}

\begin{figure}
\centering
\includegraphics[scale=0.39]{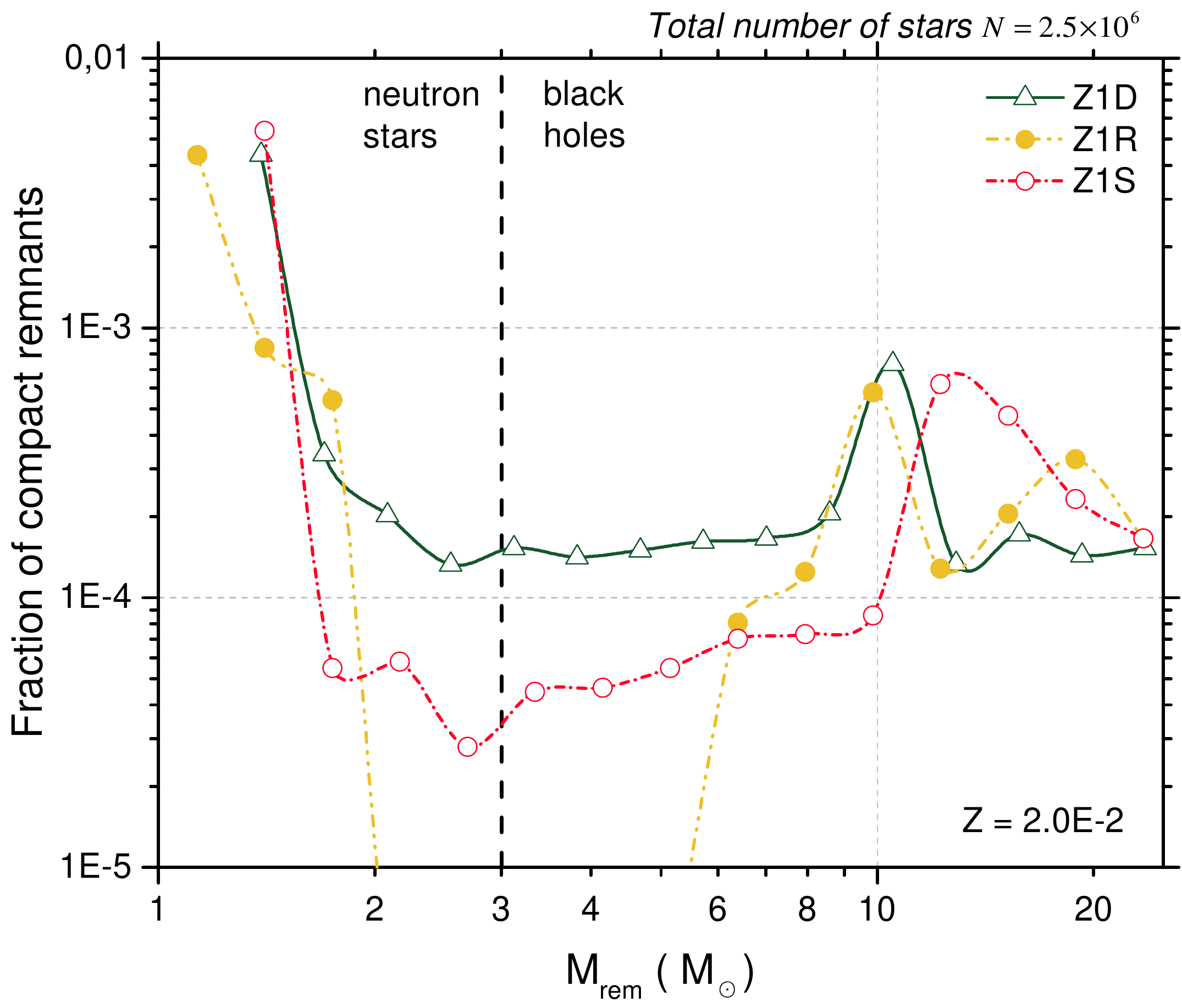}
\caption{Fraction of compact remnants, normalized to the total number of stars ($N=2.5\times{}10^6$) that initially follow a Kroupa IMF. Solid line with open triangles: \codename{} with delayed SN model (Z1D); dash-double dotted line 
with circles: \codename{} with rapid SN model (Z1R); dash-dotted line with open circles: \codename{} with 
\startrack{} recipes (Z1S). The vertical dashed line at $\mrem=3\msun$ distinguishes NSs from BHs. The curves 
have been obtained for $Z=2.0\times{10^{-2}}$.}
\label{fig:fig15}
\end{figure}

\begin{figure}
\centering
\includegraphics[scale=0.39]{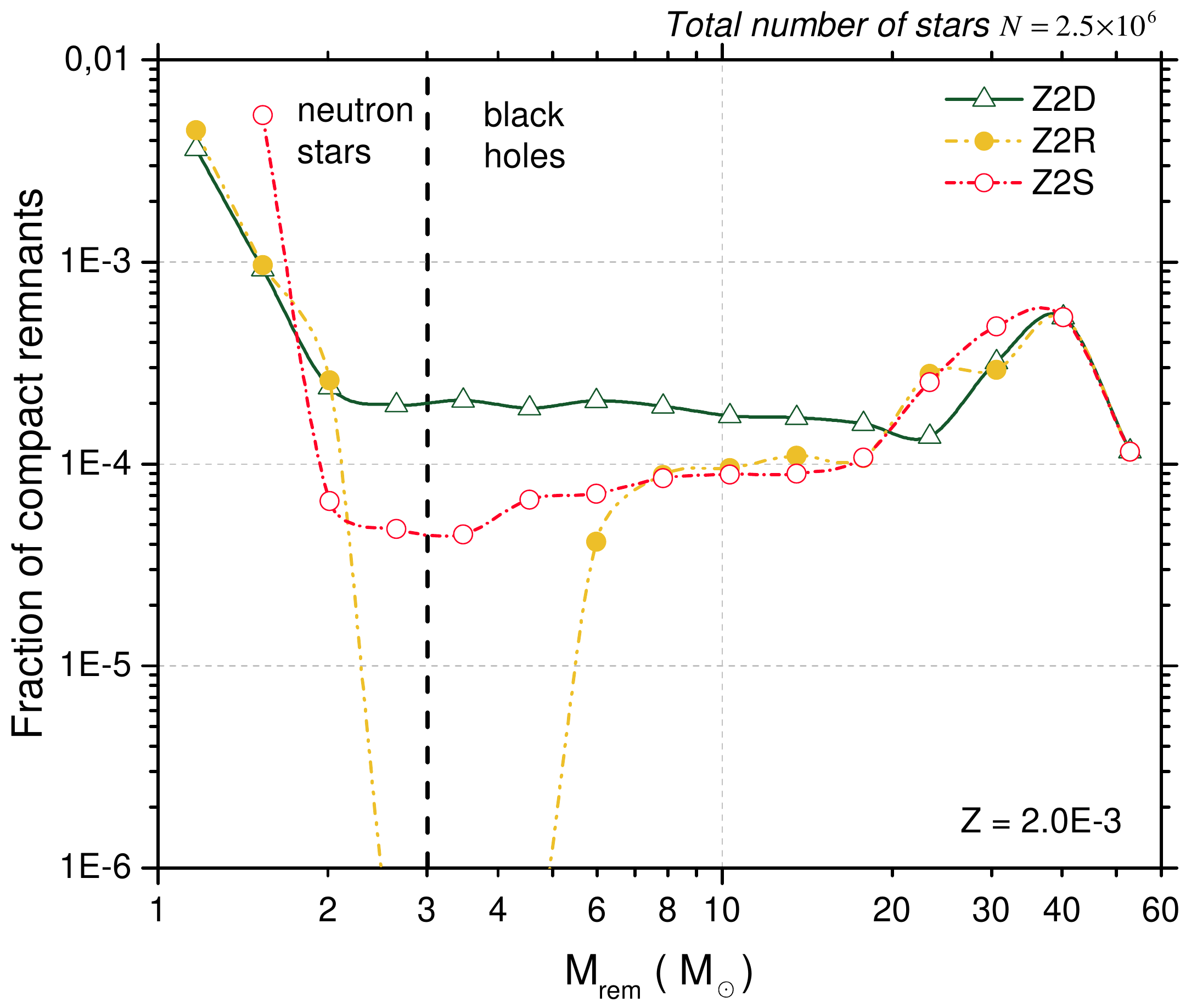}
\caption{Same as Fig. \ref{fig:fig15} but  for $Z=2.0\times{10^{-3}}$.}
\label{fig:fig16}
\end{figure}

\begin{figure}
\centering
\includegraphics[scale=0.39]{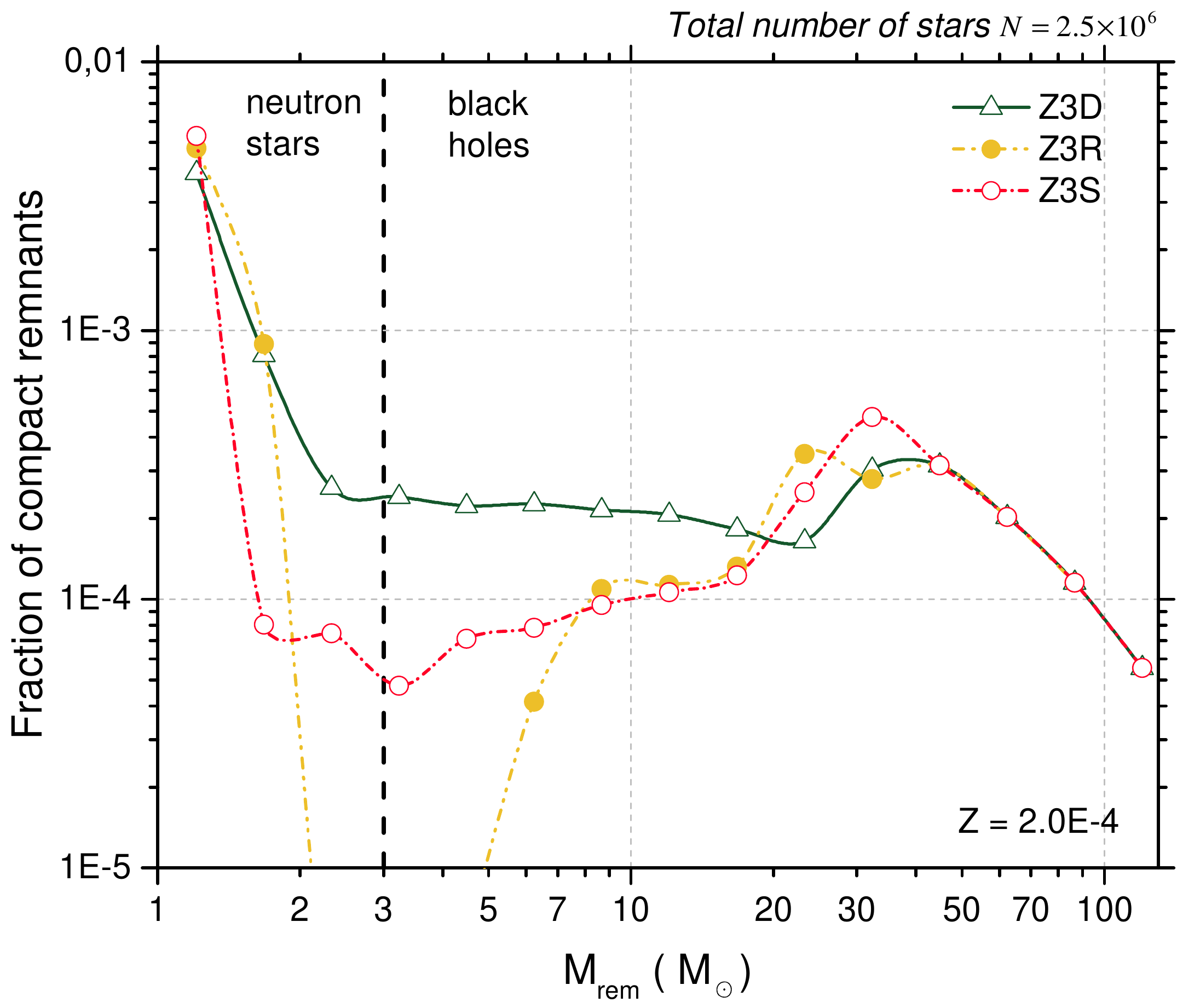}
\caption{Same as Fig. \ref{fig:fig15} but  for $Z=2.0\times{10^{-4}}$. }
\label{fig:fig17}
\end{figure}

Figure \ref{fig:fig15} shows the mass distribution of compact remnants obtained for runs~Z1D, Z1R and 
Z1S (see Table~\ref{tab:tab4}). These stellar populations have  $Z=2.0\times{10^{-2}}$ and are evolved using 
\codename{}, with the \parsec{} stellar evolution prescriptions and with different SN models (delayed SN 
model, rapid SN model and Startrack recipes for run~Z1D, Z1R and Z1S, respectively).  
Both the delayed and rapid models predict a peak of  BHs with mass 
$\sim 10\msun$ at $Z=2.0\times{10^{-2}}$,
while this peak is shifted to  $\sim 13\msun$ in the \startrack{} prescriptions. The reason for 
these peaks can be understood from Fig. \ref{fig:fig9}: for example, in the  delayed model, BHs of mass 
$9\msun\lesssim M_{\mathrm{BH}}\lesssim 11\msun$ can form from a wide range of stars (those with 
$26\msun\lesssim \mzams \lesssim 28\msun$ and with $35\msun\lesssim \mzams \lesssim 44\msun$). 

 Fig. \ref{fig:fig15} also shows that the rapid SN model predicts almost no remnants with mass between $\sim 
 2 \msun$ and $\sim 5 \msun$. This  agrees with current observations, which suggest a gap between the maximum 
 NS mass and the minimum BH mass \citep{ozel2010}\footnote{Whether the presence of this gap is physical or 
 simply due to selection biases is still 
unclear \citep{farr2011,kochanek2014,ugliano2012}, especially after the recent estimation of the BH mass of the X-ray source SWIFT J1753.5-0127, which seems to fall right into this gap \citep{neustroev2014}.}.
Figures \ref{fig:fig16} and \ref{fig:fig17} are the same as  Fig.~\ref{fig:fig15}, but for 
$Z=2.0\times{10^{-3}}$ (runs~Z2D, Z2R, Z2S) and $Z=2.0\times{10^{-4}}$ 
(runs~Z3D, Z3R, Z3S), respectively. In these Figures, the peak of BH mass distribution is at  $\sim 35- 40 
\msun$.

The mass distribution for NSs peaks at $1.3-1.6 \msun$ for all the \codename{} models, almost independently 
of metallicity. A relevant difference between the models  is that the delayed SN model 
forms a not negligible number of NSs with masses between $2\msun$ and $3\msun$ while, for the other SN 
explosion mechanisms, the vast majority of NSs have a mass below $2 \msun$.

\begin{figure}
\centering
\includegraphics[scale=0.39]{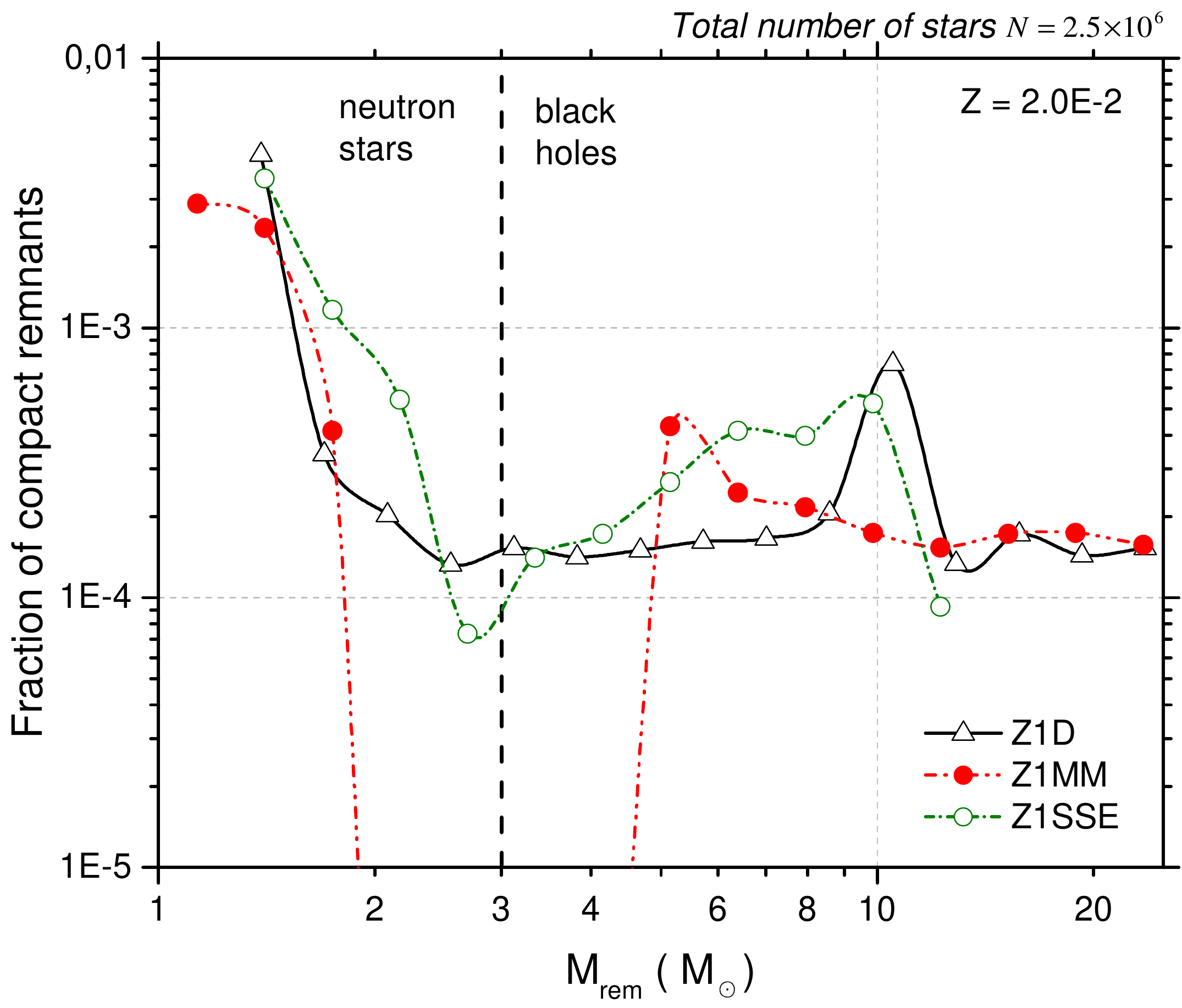}
\caption{Mass function of NSs and BHs, normalized to the total number of stars ($N=2.5\times{}10^6$) that initially follow a Kroupa IMF, obtained using three different codes: \codename{}, \starlabmm{}  and \sse{}.
 Solid line with open triangles: \codename{} with the delayed 
SN model (Z1D); dash-double dotted line with circles: \starlabmm{} (Z1MM); dash-dotted line with open 
circles: \sse{} (Z1SSE). The vertical dashed line at $\mrem=3\msun$ 
distinguishes NSs from BHs. The curves have been obtained for $Z=2.0\times{10^{-2}}$.}
\label{fig:fig18}
\end{figure}

\begin{figure}
\centering
\includegraphics[scale=0.39]{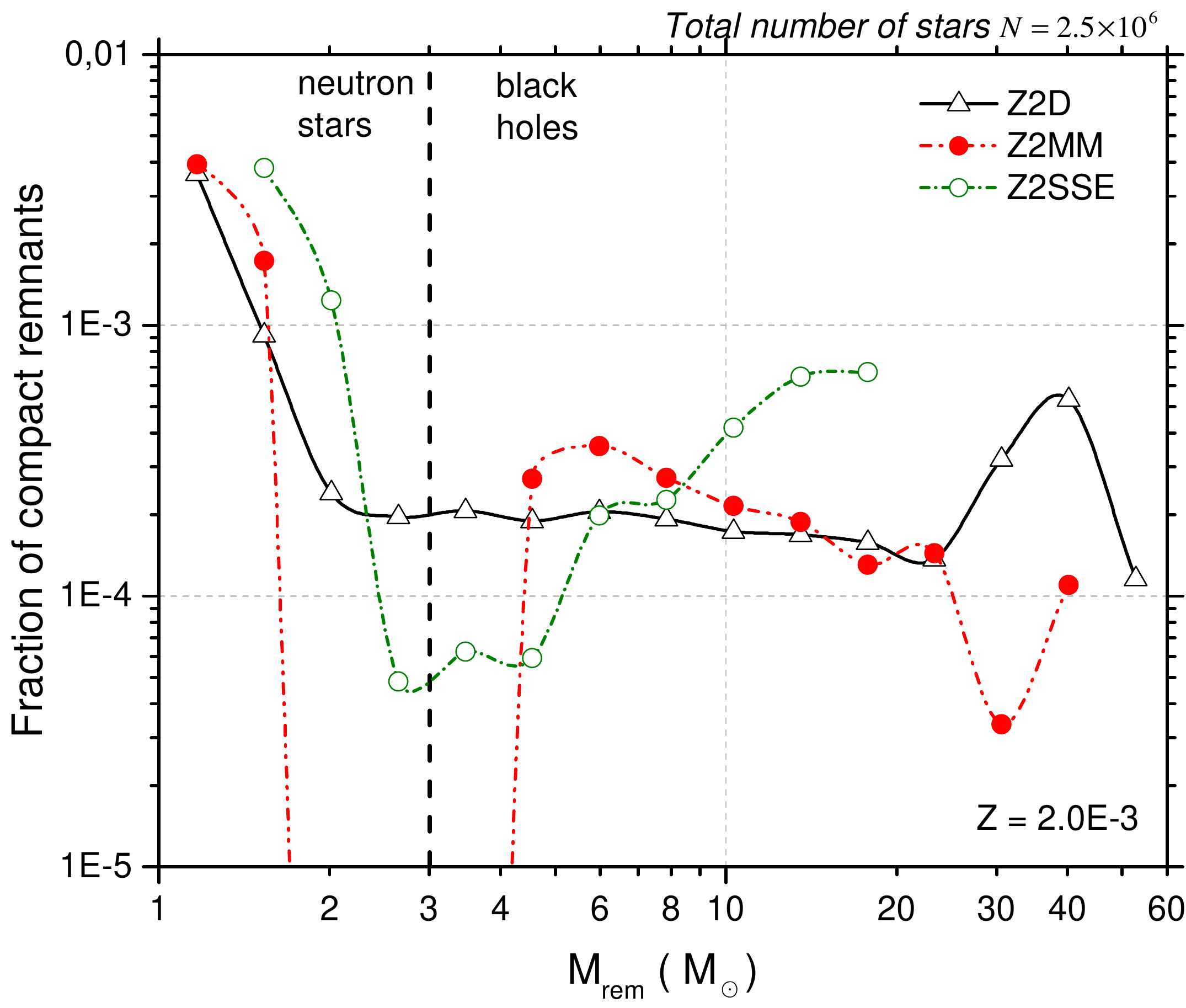}
\caption{Same as Fig. \ref{fig:fig18} but for $Z=2.0\times{10^{-3}}$.}
\label{fig:fig19}
\end{figure}

\begin{figure}
\centering
\includegraphics[scale=0.39]{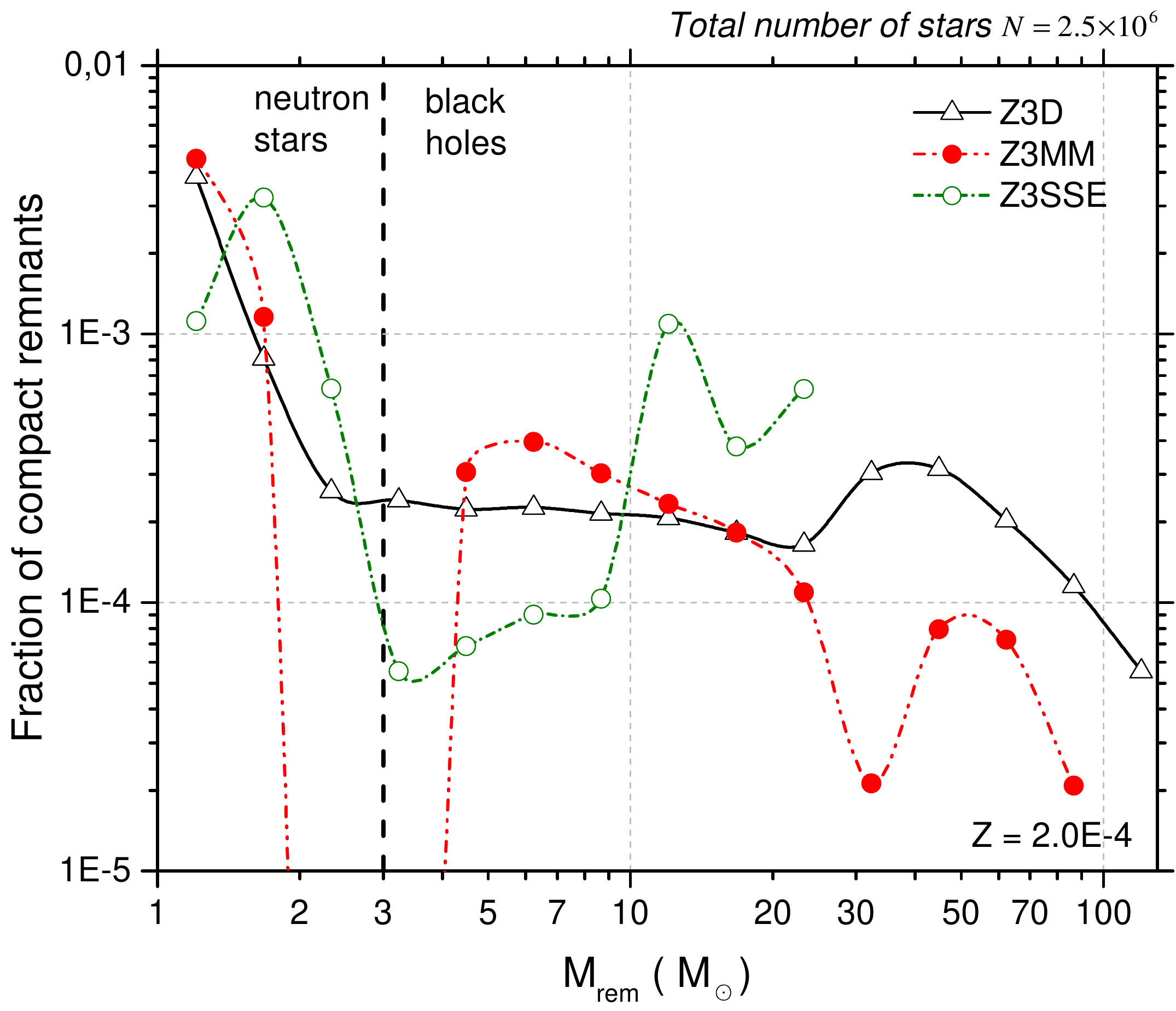}
\caption{Same as Fig. \ref{fig:fig18} but for $Z=2.0\times{10^{-4}}$.}
\label{fig:fig20}
\end{figure}

Figure \ref{fig:fig18} compares the mass distribution of compact remnants in runs~Z1D,~Z1MM 
and~Z1SSE (i.e. the same stellar population at metallicity $Z=2.0\times{10^{-2}}$, run with \codename{}, 
\starlabmm{} and \sse{}, respectively).
Run~Z1MM (\starlabmm{}) agrees with run~Z1R (\codename{} with the rapid SN mechanism) to reproduce the dearth of compact remnants with mass between $\sim 2\msun$ and $\sim 5\msun$. The majority of BHs formed in run~Z1SSE (\sse{}) have mass between $\sim 5\msun$ and $\sim 12 \msun$. \sse{} does not produce compact remnants with mass $\gtrsim 13\msun$. As to NSs, \sse{} produces more NSs with masses between $\sim 1.5 \msun$ and $2.5 \msun$ than the other codes, while \starlabmm{} does not form NSs with mass $\gtrsim 1.5 \msun$.

Figures \ref{fig:fig19} and \ref{fig:fig20} are the same as Fig. \ref{fig:fig18} but for 
$Z=2.0\times{10^{-3}}$ and $Z=2.0\times{10^{-4}}$, 
respectively. At low metallicities, \codename{} produces heavier BHs than both \starlabmm{} and \sse{}. The 
majority of BHs in both run~Z2SSE and Z3SSE have mass $\sim{} 10-20\msun$, while the BH mass in both run~Z2D  
and  Z3D peaks at about $\sim 40 \msun$. In run~Z2D  (run~Z3D) the distribution of BH masses extends up to 
$\sim 60 \msun$ ($\sim 100 \msun$).  
Tables \ref{tab:tab5} and \ref{tab:tab6} report the fraction of  BHs and massive stellar black holes (MSBHs, 
i.e. BHs with mass $>25\msun$, according to the definition by \citealt{mapelli2010}) that form in our runs.

\begin{table}
\begin{center}
\caption{Fraction of BHs, normalized to total number of stars,  obtained from \codename{}, adopting different SN explosion 
models, and from \starlabmm{} and \sse{}. D: delayed model; R: rapid model; S: \startrack{} prescriptions.}
\label{tab:tab5}
\begin{tabular*}{\columnwidth}{@{\extracolsep{\fill}} cccccc}

\toprule
Z    &  \multicolumn{3}{c}{\codename{}} & \starlabmm{}     & \sse{}    \\
\cmidrule(r){2-4}
& D & R & S & &\\
\midrule
$2.0\times{10^{-4}}$ & $2.38$ & $1.72$ & $1.94$ & $1.72$ & $2.40$ \\
$2.0\times{10^{-3}}$ & $2.40$ & $1.66$ & $1.92$ & $1.72$ & $2.28$ \\
$2.0\times{10^{-2}}$ & $2.26$ & $1.62$ & $1.86$ & $1.72$ & $2.02$ \\
\bottomrule
\end{tabular*}
\footnotesize{The values are normalized to $10^{-3}$.}\hspace*{\fill}
\end{center}
\end{table}

\begin{table}
\begin{center}
\caption{Fraction of MSBHs (that is, BHs with mass $>25\msun$), normalized to the total number of stars, obtained from \codename{}, 
adopting different SN explosion models, and from \starlabmm{} and \sse{}. D: delayed 
model; R: rapid model; S: \startrack{} prescriptions.}
\label{tab:tab6}
\begin{tabular*}{\columnwidth}{@{\extracolsep{\fill}} cccccc}

\toprule
Z    &  \multicolumn{3}{c}{\codename{}} & \starlabmm{}   & \sse{}    \\
\cmidrule(r){2-4}
& D & R & S & &\\
\midrule
$2.0\times{10^{-4}}$ & $1.04$ & $1.00$ & $1.30$ & $0.20$ & $0.16$ \\
$2.0\times{10^{-3}}$ & $1.00$ & $0.96$ & $1.24$ & $0.18$ & 0 \\
$2.0\times{10^{-2}}$ & 0 & 0 & 0 & 0 & 0 \\
\bottomrule
\end{tabular*}
\footnotesize{The values are normalized to $10^{-3}$.}\hspace*{\fill}
\end{center}
\end{table}

The fraction of BHs in Table \ref{tab:tab5} is remarkably similar in all compared codes. 
Furthermore, this number is almost independent of metallicity. On the other hand, the tested codes exhibit 
significant differences when the fraction of MSBHs is considered (Table~\ref{tab:tab6}). At metallicity 
$Z=2.0\times{10^{-2}}$, none of the compared codes form MSBHs, in agreement with the mass spectra we 
presented in Fig.~\ref{fig:fig12}. At lower metallicity, \codename{} produces, on average, $5-6$ times 
more MSBHs than \sse{} and  \starlabmm{}. Therefore, the \parsec{} stellar evolution prescriptions (combined 
with \citealt{fryer2012} SN 
models) tend to form, approximately, the same 
number of BHs as the other codes, but many more MSBHs at low metallicity.

\section{Comparison with compactness-based and two-parameter models}\label{sec:newmodels}

\begin{figure*}
	\centering
	\includegraphics[scale=0.48]{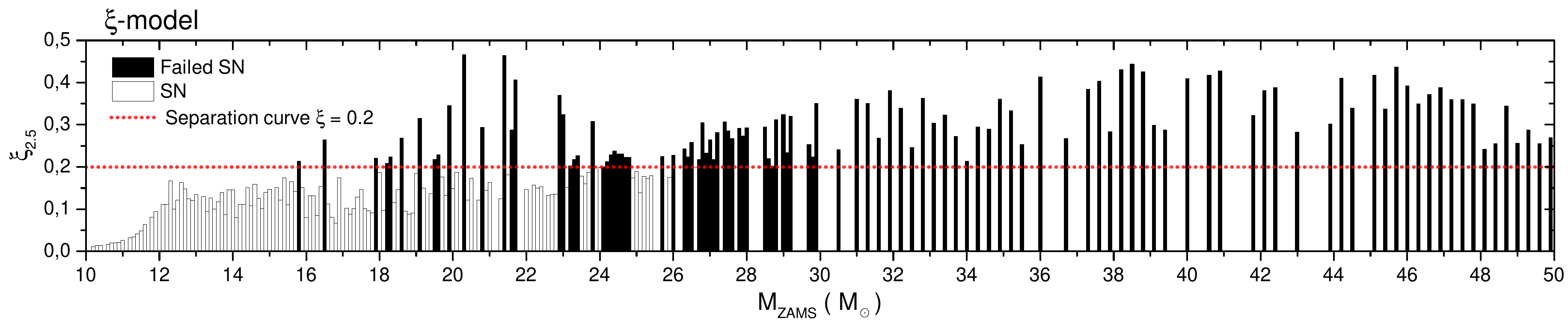}
	\caption{Value of compactness parameter at the innermost 2.5\msun{}, \comp{}, as a function of \mzams{},  
	for the \parsec{} models evolved until the Fe-core infall using \mesa{}. Black bars indicate  
	non-exploding models (failed SNe) while white bars refer to exploding models (SNe). The dotted line $\xi_{2.5}=0.2$ is the  
	threshold we chose to distinguish between SNe and failed SNe according to \citet{horiuchi2014}. The  
	simulation grid goes from $\mzams{}=10.0\msun$ up to $\mzams{}=30.0\msun$ with steps of $0.1\msun$,  and  
	from $\mzams{}=30.0\msun$ to $\mzams{}=50.0\msun$ with steps of $0.3\msun$. Some models in  the grid  are 
	not shown in the results because of numerical convergence issues.}
	\label{fig:fig21}
\end{figure*}

\begin{figure*}
	\centering
	\includegraphics[scale=0.48]{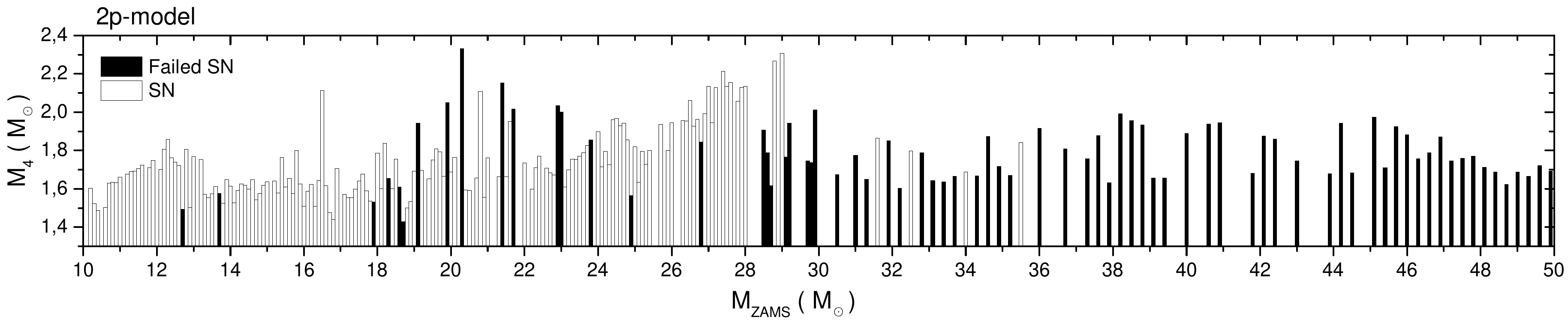}
	\caption{Same as Fig. \ref{fig:fig21} but here we show the parameter $M_4$ of the model 
	by \citet{ertl2015} as a function of \mzams{}. $M_4$ represents the baryonic mass of the proto-compact 
	object following the SN explosion event. In this case, to separate between SNe and failed SNe, 
	we used the linear function $y_{\mathrm{sep}}\left(x\right)=0.283x + 0.0430$ which corresponds to the 
	calibration w18.0 of \citet{ertl2015}.}
	\label{fig:fig22}
\end{figure*}

The \citet{fryer2012} SN models we described in Sections 3.3-3.6 (as well as the other explosion 
prescriptions adopted in 
$N$-body simulations so far) are based on a single-parameter criterion that 
discriminates between SN explosion or failed SN. In this framework, stars explode if $\mco <   
\mco{}_{\mathrm{,cut}}$ with $\mco{}_{\mathrm{,cut}} = 11.0\msun$ for the rapid and delayed models and 
$\mco{}_{\mathrm{,cut}} = 7.6\msun$ for the \startrack{} model. Recent studies have 
shown
that the link between physical properties of the progenitor star, SN properties and mass of the 
compact remnant is far from being 
trivial (\citealt{sukhbold2014,ugliano2012,oconnor2011,smartt2015,janka2012,ertl2015}). In particular, it has 
been shown that the internal structure of the stars at core collapse varies non-monotonically with \mzams{} 
(or 
\mco{}) and this may lead to different compact remnants even if the progenitors had very similar 
\mzams{}. In this Section, we highlight the main differences between criteria based on 
$\mco{}_{\mathrm{,cut}}$ and more sophisticated models, based on the structural properties of the star at the 
pre-SN stage.  

\subsection{The compactness criterion}\label{sec:compactness}
\citet{oconnor2011} suggest that the value of the compactness 
$\xi_{M}$ evaluated just outside the iron core can discriminate between SNe and failed SNe. $\xi_{M}$ is the 
ratio between the innermost mass $M$ of the star, in units of \msun{}, and the radius $R\left(M\right)$ 
containing $M$, in units of 1000 km, i.e.  
\begin{equation}
\xi_{M}\equiv\frac{M/\msun}{R\left(M\right)/1000\, \mathrm{km}}.
\end{equation}
Large values of $\xi_{M}$ favour failed SNe, while SNe occur for small values of $\xi_{M}$. 
Generally, a fiducial value of $M=2.5\msun$ is used to evaluate the compactness just outside the iron core. 
Even if the value of $\xi_{2.5}$ is sensible to changes in mass loss prescriptions and stellar evolution 
parameters (such as mixing, reaction rates, opacity, metallicity), a threshold $\xi_{2.5} \sim 0.2$ seems to 
be a reasonable value to distinguish between the occurrence of explosion and failed SNe (\citealt{horiuchi2011, smartt2015}).  Hereafter, we refer 
to the $\xi_{2.5}$-parameter model as $\xi$-model.

In order to use the $\xi$-model, \codename{} needs further information (in addition to the standard input tables described in Appendix \ref{appsec:detailsimplementation}), that is (i) the value of $R\left(M\right)$ at the core collapse stage\footnote{In our models, we identify the pre-SN stage when the collapse speed reaches $\sim 10^{8}$ cm/s.}, to evaluate \comp{} and to distinguish between SNe and failed SNe; (ii) the mass of the iron core \mfe{}, which is taken as the mass of the proto-compact object \proto{}.

Since \parsec{} numerically integrates the stellar structure up to the beginning of the CO burning phase 
only, we merged the \parsec{} wind prescriptions with the \mesa{} code (\citealt{paxton2011, paxton2013}) 
and we used \mesa{} to evolve the \parsec{} models until the iron core infall phase. Our  grid of \mesa{} simulations goes from $\mzams{}=10\msun$ up to $\mzams{}=30\msun$ with steps of $0.1\msun$, and from 
$\mzams{}=30\msun$ to $\mzams{}=50\msun$ with steps of $0.3\msun$.

Fig. \ref{fig:fig21} shows how the compactness parameter \comp{} changes as a function 
of \mzams{}. In this plot, we chose a critical compactness value of \comp{}=0.2 
(\citealt{horiuchi2011,horiuchi2014}) to separate SNe from failed SNe. The relation between \comp{} and 
\mzams{} is quite complex.
In particular, with the $\xi$-model, we distinguish at least three areas:
\begin{enumerate}
	\item range $\mzams{} \in \left[10\msun; 18\msun\right]$: the majority of stars explode as SN and  leave a NS with mass $\mfe$ (excluding fallback material);
	\item range $\mzams{} \in \left[18\msun; 26\msun\right]$:  both SNe and failed SNe occur in this mass range;
	\item range $\mzams{} > 26\msun$: the majority of stars undergo direct collapse forming a BH with mass $M_{\mathrm{BH}} = \mfin$.
\end{enumerate}
Even if the value of $\xi{}_{2.5}$ depends on many stellar evolution parameters (including the adopted mass loss recipes),  we find similar results to those obtained by other authors (e.g. \citealt{ugliano2012, ertl2015}).

\begin{figure} 
	\centering
	\includegraphics[scale=0.28]{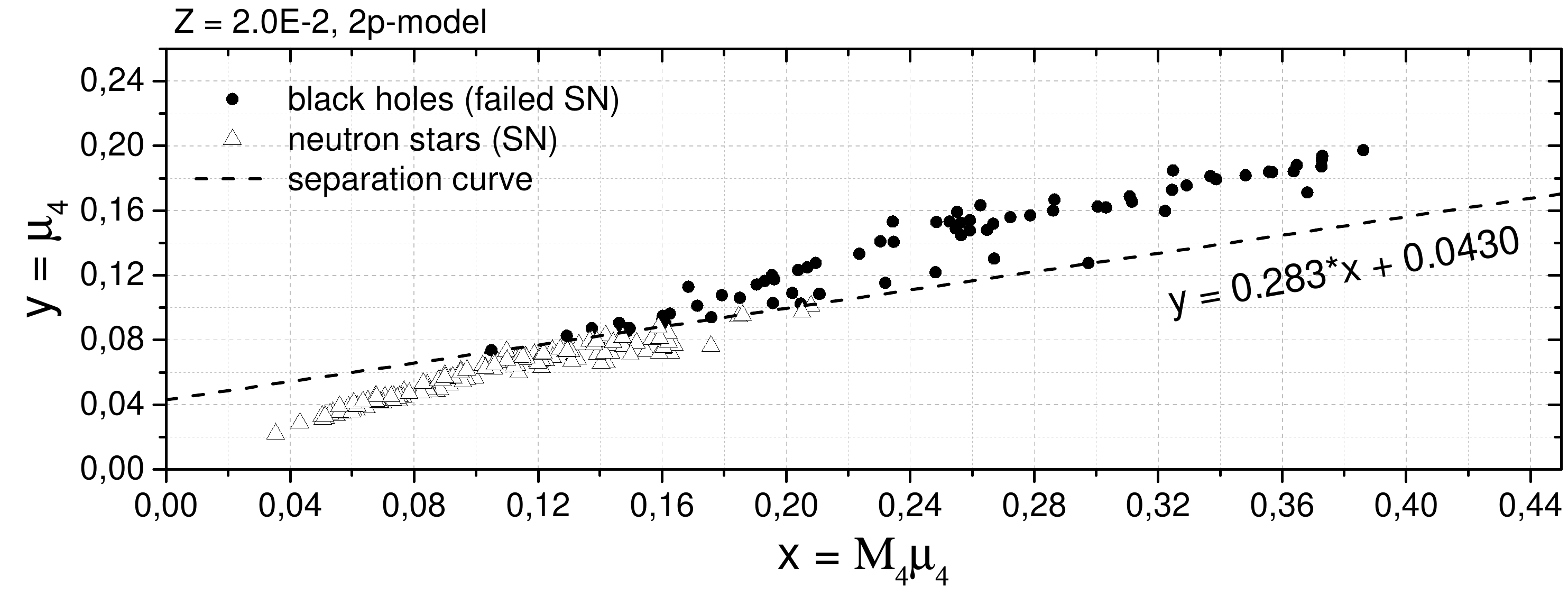}
	\caption{Representation of the results we obtained with \parsec{} progenitors in the two-parameter space 
	introduced by \citet{ertl2015}, at $Z=0.02$. Filled circles represent the formation of BHs via direct 
	collapse, while 
	open triangles refer to SNe. The dashed curve that divides SNe from failed SNe comes from calibration w18.0 of \citet{ertl2015}.}
	\label{fig:fig23}
\end{figure}


\begin{figure}
	\centering
	\includegraphics[scale=0.39]{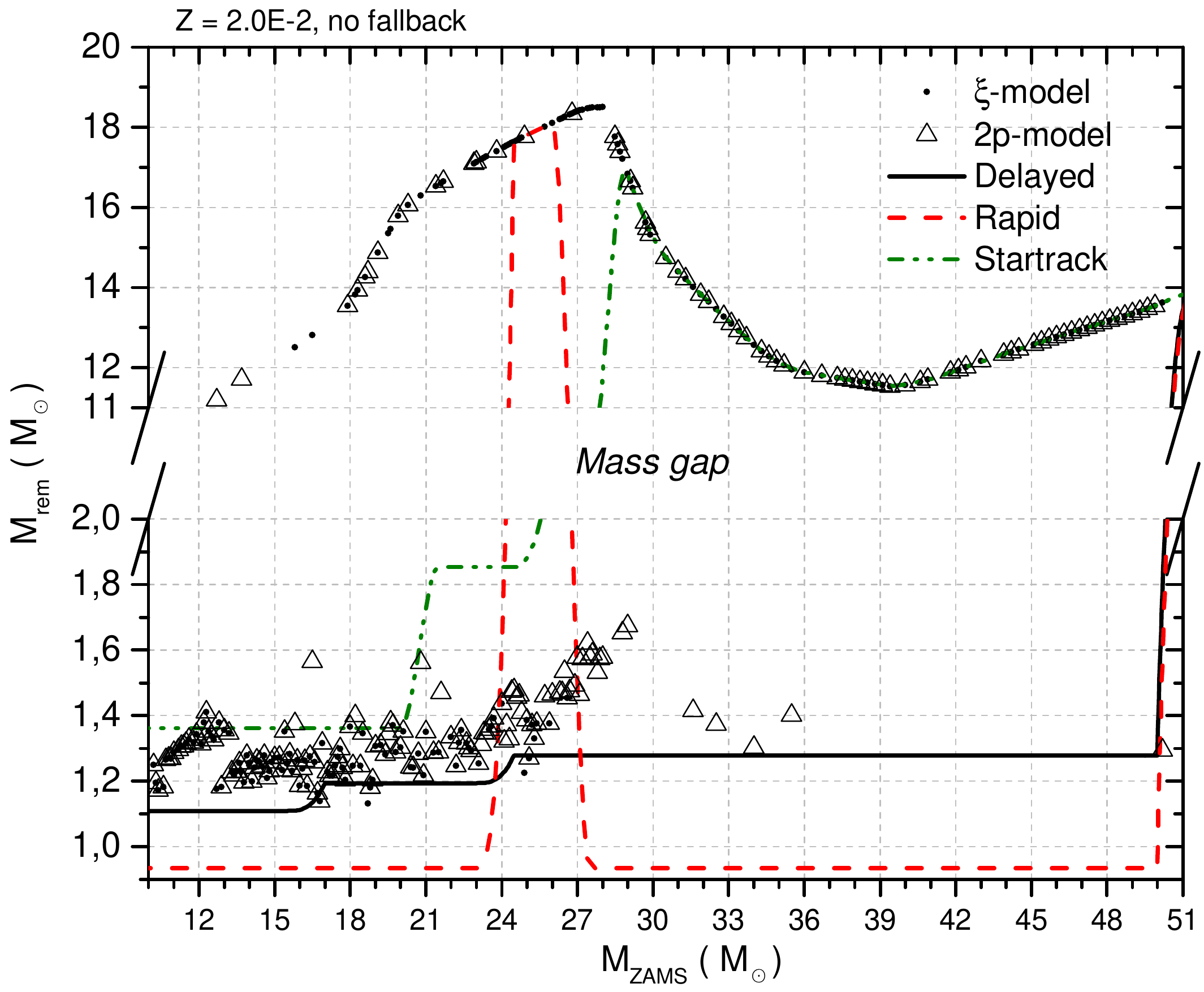}
	\caption{Mass of the compact remnant as a function of \mzams{} for  the SN explosion recipes discussed 
	in this paper, at $Z=0.02$. In particular, filled circles: $\xi{}-$model; open triangles: 2p-model; solid line: delayed SN model; long-dashed line: rapid SN model; short-dashed line: Startrack recipes. In this plot, fallback is not included. We insert a y-axis break between 
	2\msun{} and 11\msun{} to better represent the mass spectrum of both BHs and NSs for the various models 
	and the mass gap between the heaviest NS and the lightest BH.}
	\label{fig:fig24}
\end{figure}

\subsection{The two-parameter model}
 A  recent study 
by \cite{ertl2015} introduces a two-parameter criterion.
The two parameters are $M_{4}$, which represents the 
enclosed mass at a dimensionless entropy per nucleon $s=4$, and $\mu_4$, that is the mass gradient at the  
same location. Following the definition by \citet{ertl2015}, $M_{4}$ is normalized to \msun{} and $\mu_4$ is 
normalized to $10^3\,\mathrm{km}/\msun{}$. 
\cite{ertl2015} show that a separation curve exists, that divides 
exploding from non-exploding stars, in the plane $x = M_{4}\mu_4$, $y = \mu_4$. The threshold function is a straight line
\begin{equation}
\label{eq:relertl}
y_{\mathrm{sep}}\left(x\right)=k_1\,{}x + k_2
\end{equation}
where the coefficients $k_1$ and $k_2$ slightly depend on the different calibrations of the free parameters 
of \cite{ertl2015}   1D hydrodynamical simulations. Here, we use the calibration curve for the model w18.0 given by \citet{ertl2015}, 
for which $k_1=0.283$ and $k_2=0.043$. 
Progenitors with $y_{\mathrm{progenitor}} > y_{\mathrm{sep}}$ collapse directly into a BH, otherwise they 
 explode as SN. Hereafter, we refer to this model as 2p-model. In order to apply this criterion to 
 \parsec{} progenitors,  we extract the values of $M_{4}$ and $\mu_4$ from our grid of simulations run with 
 \mesa{}, coupled with the \parsec{} wind models (see Sec.~\ref{sec:compactness} for details).

Fig. \ref{fig:fig22} shows the parameter $M_{4}$ (baryonic mass of 
the remnant if the compact object is a NS) as a function of \mzams{}. Black bars indicate direct collapse 
while white bars refer to SN explosion events. In this plot, we distinguish four different regions:

\begin{enumerate}
	\item range $\mzams{} \in \left[10\msun; 18\msun\right]$: the majority of stars explode as SNe 
	and  leave a NS with baryonic mass $M_4$ (excluding fallback material);
	\item range $\mzams{} \in \left[18\msun; 24\msun\right]$:  both SNe and failed SNe occur;
	\item range $\mzams{} \in \left[24\msun; 28\msun\right]$: the majority of stars undergo SN explosion;
	\item range $\mzams{} > 28\msun$: the majority of stars form BHs through direct collapse.
\end{enumerate}

For the calibration we assume, the main difference with the $\xi$-model (see Fig. \ref{fig:fig21}) is in the 
range $\mzams{} \in \left[24\msun; 28\msun\right]$, where the 2p-model produces a significantly higher number of NSs. 
This result confirms that BHs (NSs) can form even for $\mzams{} \lesssim 25\msun$ ($\mzams{} \gtrsim 25 \msun$).

Fig. \ref{fig:fig23} shows the parameter $y \equiv \mu_4$ 
as a function of $x \equiv M_4\mu_4$, for the 2p-model. Filled circles indicate BH formation (via direct 
collapse) while open triangles refer to the production of NSs (SNe explosion). Our \parsec{} progenitors populate a narrow region in the $x-y$ parameter space, whose range is similar to that  shown in \citet{ertl2015}.

\subsection{Comparison with \citet{fryer2012} models}
Fig. \ref{fig:fig24} shows the mass spectrum of compact remnants, obtained using the $\xi$-model 
(filled circles) and the 2p-model (open triangles), as 
a function of \mzams{} (black points), at $Z=0.02$. In the same figure, we also represent the mass spectrum 
given by the delayed, rapid and \startrack{} models. Since both the $\xi$-model and the 2p-model do not 
provide prescriptions to 
evaluate the amount of mass that falls back onto the proto-compact object, all the models shown in Fig. 
\ref{fig:fig24} do not include fallback.

Overall, the mass spectrum of compact remnants resulting from either the $\xi$-model or the 2p-model  is similar to the one derived from the \startrack{} model. The main difference is that the $\xi$- and 2p-models predict a 
significant amount of BHs, due to failed SNe, for $\mzams < 30 \msun$. Using the delayed and rapid models, direct collapse occurs for $\mzams \gtrsim 50 \msun$ only.

Finally, Fig. \ref{fig:fig24} shows that there is a significant 
mass gap between the heaviest NS ($\sim 2 \msun$) and the lightest BH ($\sim 12\msun$), quite larger than the 
observed one  
(\citealt{ugliano2012,farr2011,ozel2010,neustroev2014,kochanek2014}). Still, the area between 2 
\msun{} and 12 \msun{} may be populated by  NSs that accrete mass through the fallback mechanism\footnote{\citet{ertl2015} show that a typical amount of mass that falls back onto the proto-compact 
object is $\sim 0.05 \msun$. Values larger than 1 \msun are rare (only 6 events over $\sim 600$ progenitor 
models).}, and/or by NSs that accrete mass from a companion, in a binary system.

\section{Conclusions}\label{sec:conclusions}
The mass spectrum of BHs is still an open issue: only a few dynamical mass measurements of BHs are available 
\citep{ozel2010}, while theoretical models are affected by the uncertainties on SN explosion and massive star 
evolution. In this paper, we derive the mass spectrum of compact remnants based on the new stellar evolution 
models implemented in \parsec{} \citep{bressan2012,chen2014,tang2014}, combined with different recipes for SN 
explosion: the rapid and delayed SN models presented in \cite{fryer2012}, the SN model implemented in the 
\startrack{} code \citep{belc2008}, the SN recipes included in \starlab{} through the \seba{} module 
\citep{zwart2001},  and (for $Z=0.02$) the $\xi{}-$ and the 2p-models \citep{oconnor2011,ertl2015}.

These recipes for stellar evolution and SN explosion are implemented in our new public tool \codename{}, 
which can be used both as a stand-alone population-synthesis code or as a module in several N-body codes 
(\starlab{}, \higpus{}).  \codename{} is extremely versatile, because it calculates the mass, radius, 
luminosity, temperature and chemical evolution of a star based on stellar-evolution tables. We adopt 
stellar-evolution tables that have been generated with the \parsec{} code, but these can be substituted with 
different stellar-evolution models in a fast and simple way.

With respect to previous stellar-evolution codes, \parsec{} predicts significantly larger values of \mfin{} 
and \mco{} at low metallicity ($\lesssim{}2\times{}10^{-3}$, Figures~\ref{fig:fig4} and \ref{fig:fig5}). We 
find differences up to $\sim{}80$ \% between the value of  \mfin{} calculated by \parsec{} and the fitting 
formulas implemented in \sse{} (Fig.~\ref{fig:fig3}). This implies that \codename{} predicts substantially larger BH 
masses at low metallicity, since the mass of the compact remnants  depends on \mfin{} and \mco{} in the SN models developed by \citet{fryer2012}.

Moreover, for a metallicity $Z=0.02$ and for $\mzams \leq 50\msun$, we also present the mass 
spectrum of NSs and BHs given by the $\xi{}$-model \citep{oconnor2011} and the 2p-model \citep{ertl2015}. 
These models depend on stellar structural parameters evaluated at the time of iron core infall. Coupling these new prescriptions with the \parsec{} 
stellar models, we find that the relation between progenitor mass and remnant mass is quite complex, 
especially in the range $\mzams \in \left[18\msun;30\msun\right]$ 
(see Figures \ref{fig:fig21} and \ref{fig:fig22}). A detailed study that 
considers also $Z\neq 0.02$ and $\mzams > 50\msun$ is still in progress.

Using the \citet{fryer2012} models, 
we find that the maximum BH mass found with \codename{} is $\sim{}25$, 60 
and 130 M$_\odot$ at 
$Z=2\times{}10^{-2}$, 
$2\times{}10^{-3}$ and $2\times{}10^{-4}$, respectively. 
Mass loss by stellar winds plays a major role in determining the mass of BHs for very massive stars 
($\gtrsim{}90$ M$_\odot{}$), almost independently of the adopted SN recipe. In contrast, the adopted SN model is very important for lower BH masses, and for 
the 
transition between NSs and BHs (Figures \ref{fig:fig8}, \ref{fig:fig9}, 
\ref{fig:fig10} and \ref{fig:fig11}): according to the delayed SN  model, stars with 
$M_{\rm ZAMS}>19$ M$_\odot$   end their life as BHs, while  this limit is $M_{\rm ZAMS}>24-25$ M$_\odot$ if 
the rapid SN mechanism or the \seba{} recipes are assumed.

As a consequence, the rapid SN mechanism and the recipes implemented in \seba{} predict a gap between the 
maximum mass of NSs and the minimum mass of BHs, while the delayed SN model (and the \startrack{} recipes) 
suggest a smooth transition between NSs and BHs (Figs.~\ref{fig:fig15},~\ref{fig:fig16} and 
~\ref{fig:fig17}).  The distribution of dynamically measured BH and NS masses in the local Universe suggests 
the existence of a gap between NS and BH masses \citep{ozel2010}, even if the statistical significance of 
this result is still debated (\citealt{farr2011,ugliano2012,kochanek2014,neustroev2014}).

According to \codename{} (with either the delayed or the rapid SN model), at $Z=2\times{}10^{-2}$ most BHs 
have mass $8-12$ M$_\odot$, while at $2\times{}10^{-3}\ge{}Z\ge{}2\times{}10^{-4}$  most BHs have mass 
$20-60$  M$_\odot$ (Figs.~\ref{fig:fig15},~\ref{fig:fig16} and ~\ref{fig:fig17}).

For a stellar population following the Kroupa IMF, the total number of BHs predicted by \codename{} in its 
various SN flavours is remarkably similar to other codes, such as \starlabmm{} \citep{mapelli2013} and \sse{} 
\citep{hurley2000}. Furthermore, the fraction of BHs is almost independent of metallicity.
 On the other hand,  the fraction of MSBHs (i.e. BHs with mass $>25$ M$_\odot$) strongly depends on the 
 metallicity and on the assumed stellar evolution recipes. At metallicity $Z=2.0\times{10^{-2}}$, no MSBHs 
 form from single-star evolution, using either \codename{} or \starlabmm{} or \sse{}. At lower metallicity, 
 \codename{} produces, on average, $5-6$ times more MSBHs than \sse{} and  \starlabmm{}.

This might have dramatic consequences for both the number of X-ray binaries powered by MSBHs and the 
detection of gravitational waves by BH-BH binary mergers. As to X-ray binaries, models by \cite{mapelli2014}, 
based on \starlabmm{}, indicate that MSBHs are expected to power $\sim{}20$ \% of the Roche-lobe overflow BH 
binaries in a young star cluster with $Z\lesssim{}2\times{}10^{-3}$. With the recipes implemented in 
\codename{}, the fraction of X-ray binaries powered by MSBHs might be substantially higher. 
On the other hand, quantifying the difference with previous studies is non trivial, because the evolution of 
binary systems and  dynamical encounters in star clusters can significantly affect the demographics of BH 
binaries. In a forthcoming study, we will use \codename{} to investigate the demographics of X-ray binaries 
and BH-BH binaries in star clusters.

\section*{Acknowledgements}
We thank the anonymous referee for their suggestions that helped us to improve the 
manuscript.
MM and MS acknowledge financial support from the Italian Ministry of Education, University and Research 
(MIUR) through grant FIRB 2012 RBFR12PM1F. The authors acknowledge financial support from INAF through grant 
PRIN-2014-14. 

\bibliographystyle{mn2e}
\bibliography{spera_mapelli_bressan}

\appendix

\section{Notes on the  implementation of \codename{}}
\label{appsec:detailsimplementation}
\subsection{General scheme}
 As we discussed in Sec. \ref{subsec:stevoimpl}, \codename{} can be used as a stand-alone population 
 synthesis code and/or can be easily linked to several \nbody{} codes. It is extremely versatile because it 
 relies upon a set of isochrones as input files; this means that we can easily change stellar evolution 
 recipes by simply substituting the input tables.

\codename{} reads a single file that contains a set of isochrones, which are provided by a stellar evolution 
code (by \parsec{}, in the current version of \codename{}). By default, the name of this file must have the 
form  \texttt{tableZn.dat}, where \texttt{n} indicates the metallicity $Z$. Inside this file, the isochrones 
are separated by a line reporting the age (in Gyr), and the number of points of the following isochrone. Each 
isochrone is composed of 9 columns that indicate (1) the initial mass of the star, (2) its present mass, (3) 
the logarithm of luminosity, (4) the effective temperature, (5) the logarithm of radius and (6) the logarithm 
of surface gravity, (7) the Helium and (8) Carbon-Oxygen core mass and (9) the stellar type at the current 
age. All given values are in solar units, except for the effective temperature, which is absolute and 
expressed in Kelvin. We stress that the isochrones do not need to be equally spaced in mass or other 
quantities.

In order to speed up the calculations, \codename{} reads the isochrone file and rearranges it in a more convenient way. First of all, an equally spaced grid of masses is chosen\footnote{By default, the grid goes from $0.1\msun$ to $150\msun$ with steps of $0.5\msun$.}. For each star in the grid, we construct the time evolution of its physical parameters, recording information whenever the value of a generic stellar parameter is varied by more then $5\%$. The result is stored in 7 different files containing the time evolution of masses, radii, luminosities, stellar phases, Carbon-Oxygen core mass, Helium core mass and the corresponding ages when the stellar parameters need to be updated.

These 7 files are then loaded in a 3-dimensional structure where the first index (line number, $L$) identifies the initial mass of the star. The second index (column number, $C$) gives information about the current stellar age and the third index, $P$, refers to the specific stellar parameter we need to read or write. Thus, $L$ ranges between 1 and the number of points of the grid of masses, $1\leq P \leq 7$, and $C$ varies from $1$ to the number of update points needed for a generic star.

At the beginning of the integration, it is possible to associate two different mass indexes, $L_1$ and $L_2$, 
to each star in order to uniquely identify its position in the grid. For example, let us consider a grid of 
masses that goes from $0.1\msun$ to $150\msun$ with steps of $0.5\msun$.  The evolution of a star $S$ of mass 
$M_s=50.3$\msun will be derived interpolating the evolutionary tracks of the nearest neighbour stars, that is 
$M_1=50$\msun and $M_2=50.5$\msun, and we can compute the stellar parameters of the star $S$ using the weights

\begin{equation}
\begin{split}
\alpha_1&=\frac{M_2-M_s}{\left(M_s-M_1\right)+\left(M_2-M_s\right)}\\
\alpha_2&=\frac{M_s-M_1}{\left(M_s-M_1\right)+\left(M_2-M_s\right)}.
\end{split}
\label{eq:alphacoeff}
\end{equation}

To evolve the parameters of a generic star, we use linear interpolations. Let us consider again a test star $S$ with initial mass $M_s\left(t=0\right)=50.3\msun$. At time $t=t_1$ , this star will have a mass $M_s\left(t_1\right)$. In order to evolve the star at time $t_2 = t_1 + \Delta t$, we need to use the information of its neighbour grid stars of mass $M_1\left(0\right)=50$\msun and $M_2\left(0\right)=50.5$\msun. First of all, the code must compute the quantities $M_1\left(t_2\right)$ and $M_2\left(t_2\right)$. In general, a generic time $t_2$ will not be included in the tables. Thus, \codename{} reads the tables and searches  the values $M_1\left(t_3\right)$, $M_1\left(t_4\right)$, $M_2\left(t_5\right)$ and $M_2\left(t_6\right)$ such that  $t_3 \lesssim t_2 \lesssim t_4$ and $t_5 \lesssim t_2 \lesssim t_6$. The code then calculates $M_1\left(t_2\right)$ and $M_2\left(t_2\right)$ with a linear interpolation:
\begin{equation}
\begin{split}
M_1\left(t_2\right) &= m_1t_2+q_1\\
M_2\left(t_2\right) &= m_2t_2+q_2\\
\end{split}
\end{equation}
where
\begin{equation}
\begin{split}
m_1 &= \frac{M_1\left(t_4\right) - M_1\left(t_3\right)}{t_4-t_3}\\
m_2 &= \frac{M_2\left(t_6\right) - M_2\left(t_5\right)}{t_6-t_5}\\
q_1 &= M_1\left(t_4\right) - m_1t_4\\
q_2 &= M_2\left(t_6\right) - m_2t_6.
\end{split}
\end{equation}

Finally, the value $M_s\left(t_2\right)$ is derived with a further linear interpolation, that is

\begin{equation}
M_s\left(t_1\right) = \alpha_1 M_1\left(t_2\right) + \alpha_2 M_2\left(t_2\right)
\end{equation}

with weights $\alpha_1$ and $\alpha_2$ given in equation~\ref{eq:alphacoeff}. The same procedure is adopted to
obtain the other stellar parameters needed at a given age.

\subsection{Integration of \codename{} in \starlab{}}

As discussed in Section \ref{subsec:stevoimpl}, we have merged \codename{} with the \starlab{} software 
environment \citep{zwart2001}, and with the direct \nbody{} code \higpus{} (\cite{dolcetta2013}; Spera, in 
preparation). In order to combine \codename{} with \starlab{}, we modified the \seba{} stellar evolution 
module. In particular, \seba{} is a C++ module based on a structure of classes, in which each
class approximately corresponds to a stellar evolution phase. In order to easily match the \seba{} internal
organization and the implemented transitions between stellar evolution phases, we identify the main stellar
evolution phases using integer indexes. Namely,
\begin{itemize}
	\item \textbf{0} identifies pre-MS and MS stars that are mapped to the \seba{} class
	\texttt{main\_sequence};
	\item \textbf{1} indicates stars in the sub-giant phase; in this case, we have a one to one correspondence
	with the \texttt{sub\_giant} class;
	\item \textbf{2} groups several categories of stars (among which, red giants, blue and red super giants,
	LBVs and WRs) in the \texttt{hyper\_giant} class;
	\item \textbf{3} refers to core helium burning stars, collected in the \texttt{horizontal\_branch} class;
	\item \textbf{4} corresponds to stars in the early asymptotic giant branch (E-AGB) phase, mapped to the
	\texttt{hyper\_giant} class.
\end{itemize}

Using our simplified scheme, we lose information about some specific characteristics of the stars during the
numerical integration; for instance, we do not know if a star is a WR, a LBV or a blue super giant or a red super giant. Anyway, all these features can be recovered {\it a
posteriori} by the ages, radii, luminosities and temperatures printed in the output files.

During the pre-MS and MS phases we evolve mass, luminosity and radius of the stars
following our input tables by means of linear interpolations in time and mass. Stellar evolution continues until the function \texttt{create\_remnant()} is called. This routine contains our updated recipes for SN
explosion, and converts the star into a compact remnant, which can be either a WD, a NS or a BH,
depending on the final state of the star.

At present, since \parsec{} does not include evolutionary prescriptions for stars that undergo the thermally-pulsing AGB phase (TP-AGB), we use the \seba{} built-in class \texttt{super\_giant} to follow their evolution through this stage (see \citet{zwart2001} for the details). In particular, we assume that the stars that undergo the TP-AGB phase are those with $\mzams \lesssim M_{\mathrm{up}}=7\msun$.

\section{SN explosion mechanisms in \codename{}}
\label{appsec:summaryfryer}
Here we summarize the main 
features of the \citet{fryer2012} recipes. 
\subsubsection{\startrack{} model}
In the case of \startrack{} recipes, stars form a proto-compact object of mass \proto{} given by

\begin{equation}
\proto=
\begin{cases}
1.50 \msun &  \mco<4.82\msun \\
2.11 \msun &  4.82\msun\leq\mco\ <6.31\msun \\
0.69 \mco - 2.26\msun &  6.31\msun \leq \mco < 6.75\msun \\
0.37 \mco - 0.07\msun &  \mco \geq 6.75 \msun.
\end{cases}
\end{equation}
\ffb{} is the {\em fractional fallback parameter}, and is such that $\mfb=\ffb\left(\mfin-\proto\right)$, 
where \mfin{} is the final
mass of the star. According to \startrack{} prescriptions, the values of \ffb{} are the following:

\begin{equation}
\ffb=
\begin{cases}
0 &  \mco<5.0\msun \\
0.378\mco-1.889 &  5.0\msun\leq\mco <7.6\msun \\
1.0 &  \mco \geq 7.6 \msun .
\end{cases}
\label{eq:startrack}
\end{equation}
From the baryonic mass of the remnant $M_{rem,bar}=\proto+\mfb$, we can obtain its gravitational mass
$M_{rem,grav}$ taking into account neutrino losses. When $\mco{}\geq{}7.6$ M$_\odot$, the \startrack{} 
recipes request $\ffb=1$ (in eq. \ref{eq:startrack}), i.e. the entire final mass of the star goes into the 
remnant mass. This means that the direct collapse of a star into a BH occurs if $\mco{}\geq{}7.6$ M$_\odot$, 
according to \startrack{} prescriptions.

For NSs we use the expression given
by \citet{timmes1996}, for which

\begin{equation}
\label{eq:bar2gravNS}
M_{rem,grav}=\frac{\sqrt{1+0.3M_{rem,bar}}-1}{0.15}.
\end{equation}
For BHs we use the formula

\begin{equation}
M_{rem,grav}=0.9M_{rem,bar},
\label{eq:remgrav}
\end{equation}
following the approach described in \citet{fryer2012}.

\subsubsection{Rapid SN model}
For the rapid SN mechanism, a fixed mass of the proto-compact object, $\proto=1.0\msun$, is assumed.
In this case, the coefficient \ffb{}
is given by

\begin{equation}
\ffb=
\begin{cases}
\dfrac{0.2}{\mfin-\proto} &  \mco<2.5\msun \\
\dfrac{0.286\mco-0.514}{\mfin-\proto} &  2.5\msun\leq\mco <6.0\msun \\
1.0 & 6.0\msun \leq \mco < 7.0\msun \\
\alpha_R\mco+ \beta_R &  7.0\msun\leq \mco <11.0\msun \\
1.0 & \mco \geq 11.0\msun
\label{eq:rapid}
\end{cases}
\end{equation}where

\begin{equation}
\begin{split}
\alpha_R&\equiv 0.25 - \frac{1.275}{\mfin-\proto}\\
\beta_R&\equiv 1-11\alpha_R .
\end{split}
\end{equation}

This means that the direct collapse of a star into a BH occurs if $6.0\msun \leq \mco \leq 7.0\msun$ and if 
$\mco{}\geq{}11$ M$_\odot$ (equation~\ref{eq:rapid}), according to the rapid SN model.

\subsubsection{Delayed SN model}
For the delayed SN mechanism, the prescriptions for the mass of the proto-compact object are

\begin{equation}
\proto=
\begin{cases}
1.2 \msun &  \mco<3.5\msun \\
1.3 \msun &  3.5\msun\leq\mco <6.0\msun \\
1.4 \msun &  6.0\msun\leq\mco <11.0\msun \\
1.6 \msun &  \mco \geq 11.0 \msun.
\end{cases}
\end{equation}
The amount of fallback is determined using the following relations

\begin{equation}
\ffb=
\begin{cases}
\dfrac{0.2}{\mfin-\proto} &  \mco<2.5\msun \\
\dfrac{0.5\mco-1.05\msun}{\mfin-\proto} &  2.5\msun\leq\mco <3.5\msun \\
\alpha_D\mco+ \beta_D &  3.5\msun\leq\mco <11.0\msun \\
1.0 & \mco \geq 11.0\msun
\label{eq:delayed}
\end{cases}
\end{equation}
where

\begin{equation}
\begin{split}
\alpha_D&\equiv 0.133 - \frac{0.093}{\mfin-\proto}\\
\beta_D&\equiv 1-11\alpha_D .
\end{split}
\end{equation}

Thus, the direct collapse of a star into a BH occurs if $\mco{}\geq{}11$ M$_\odot$ 
(equation~\ref{eq:delayed}), according to the delayed SN model (i.e. the same as the rapid SN model, but 
significantly larger than in the \startrack{} recipes).

\section{General fitting formula for \mrem{}}
\label{appsec:fittingformula}
We report a fitting formula that express the compact remnant mass \mrem{} as a function of \mzams{} and  $Z$. 
The following formula has been obtained by fitting the outputs of \codename{} with the delayed SN model and 
the \parsec{} stellar evolution isochrones. The value of \mrem{} obtained with the fitting formula deviate 
from the outputs of \codename{} by $\lesssim{}10$~\%.

First, we express \mrem{} as a function of \mco{} and $Z$ (from  Fig. \ref{fig:fig7}).
For $Z\leq 5.0\times{10^{-4}}$, the best fitting curve for \mrem{} is given by:

\begin{equation}
\mrem=
\begin{cases}
{\rm max}\left(p\left(\mco\right), 1.27\msun\right) \\
 \text{\hspace{15pt} if } \mco\leq5\msun \\
 \\
p\left(\mco\right) \\
 \text{\hspace{15pt} if } 5\msun < \mco < 10\msun \\
 \\
{\rm min}\left(p\left(\mco\right), f\left(\mco,Z\right)\right) \\
 \text{\hspace{15pt} if } \mco \geq 10\msun,
\end{cases}
\label{eq:fitmremvsmco_lowz}
\end{equation}
where
\begin{equation}
\begin{split}
p\left(\mco\right) &= -2.333 + 0.1559\mco + 0.2700\mco^2\\
f\left(\mco,Z\right) &= m\left(Z\right)\mco + q\left(Z\right)
\end{split}
\end{equation}
with coefficients
\begin{equation}
\begin{split}
m\left(Z\right)&= -6.476\times{10^2}Z+1.911\\
q\left(Z\right)&= 2.300\times{10^3}Z+11.67.
\end{split}
\end{equation}

For $Z > 5.0\times{10^{-4}}$ we have
\begin{equation}
\mrem=
\begin{cases}
{\rm max}\left(h\left(\mco,Z\right), 1.27\msun\right) \\
 \text{\hspace{15pt} if } \mco\leq5\msun \\
 \\
h\left(\mco,Z\right) \\
 \text{\hspace{15pt} if } 5\msun < \mco < 10\msun \\
 \\
{\rm max}\left(h\left(\mco,Z\right), f\left(\mco,Z\right)\right) \\
 \text{\hspace{15pt} if } \mco \geq 10\msun,
\end{cases}
\label{eq:fitmremvsmco_highz}
\end{equation}

where
\begin{equation}
\begin{split}
h\left(\mco,Z\right) &= A_1\left(Z\right) +
\dfrac{A_2\left(Z\right)-A_1\left(Z\right)}{1+10^{\left(L\left(Z\right)-\mco\right)\eta\left(Z\right)}}\\
f\left(\mco,Z\right) &= m\left(Z\right)\mco + q\left(Z\right).
\end{split}
\end{equation}

For $Z \geq 1.0\times{10^{-3}}$, the coefficients of the function $h\left(\mco,Z\right)$ are
\begin{equation}
\begin{split}
A_1\left(Z\right) &= 1.340 - \dfrac{29.46}{1+\left(\dfrac{Z}{1.110\times{10^{-3}}}\right)^{2.361}}\\
A_2\left(Z\right) &= 80.22 - 74.73 \dfrac{Z^{0.965}}{2.720\times{10^{-3}}+Z^{0.965}}\\
L\left(Z\right) &= 5.683 + \dfrac{3.533}{1+\left(\dfrac{Z}{7.430\times{10^{-3}}}\right)^{1.993}}\\
\eta\left(Z\right) &= 1.066 - \dfrac{1.121}{1+\left(\dfrac{Z}{2.558\times{10^{-2}}}\right)^{0.609}},
\end{split}
\end{equation}
while for $Z < 1.0\times{10^{-3}}$, the coefficients of the function $h\left(\mco,Z\right)$ are
\begin{equation}
\begin{split}
A_1\left(Z\right) &= 1.105\times{10^5}Z-1.258\times{10^2}\\
A_2\left(Z\right) &= 91.56 - 1.957\times{10^4}Z - 1.558\times{10^{7}}Z^2\\
L\left(Z\right) &= 1.134\times{10^4}Z - 2.143\\
\eta\left(Z\right) &= 3.090\times{10^{-2}} - 22.30Z + 7.363\times{10^{4}}Z^2.
\end{split}
\end{equation}

For $Z \geq 2.0\times{10^{-3}}$, the coefficients of the function $f\left(\mco,Z\right)$ are independent of
$Z$,

\begin{equation}
\begin{split}
m &= 1.217\\
q &= 1.061,
\end{split}
\end{equation}
while, for $1.0 \times{10^{-3}} \leq Z < 2.0\times{10^{-3}}$ we have

\begin{equation}
\begin{split}
m &= -43.82Z + 1.304\\
q &= -1.296\times{10^{4}}Z + 26.98,
\end{split}
\end{equation}
and for $Z < 1.0 \times{10^{-3}}$

\begin{equation}
\begin{split}
m &= -6.476\times{}10^2Z+1.911\\
q &= 2.300\times{10^3}Z+11.67.
\end{split}
\end{equation}

Furthermore, \mco{} can be expressed as a function of \mzams{} and $Z$, by
fitting the curves of Fig. \ref{fig:fig5}. The functional form of the fit is

\begin{equation}
\begin{split}
\mco &= -2.0 +\left[B_1\left(Z\right)+2.0\right]\left[g\left(Z, \mzams;K_1,\delta_1\right)\right. + \\
 + &  \left.g\left(Z, \mzams;K_2,\delta_2\right)\right],
\end{split}
\label{eq:fitmcovsmzams}
\end{equation}
where

\begin{equation}
g\left(Z,\mzams;x,y\right) \equiv \dfrac{0.5}{1+10^{\left(x\left(Z\right)-\mzams\right)y\left(Z\right)}}.
\end{equation}

For $Z > 4.0\times{10^{-3}}$ the coefficients are

\begin{equation}
\begin{split}
B_1\left(Z\right) &= 59.63-2.969\times{10^3}Z+4.988\times{10^4}Z^2 \\
K_1\left(Z\right) &= 45.04-2.176\times{10^3}Z+3.806\times{10^4}Z^2\\
K_2\left(Z\right) &= 1.389\times{10^2}-4.664\times{10^3}Z+5.106\times{10^4}Z^2\\
\delta_1\left(Z\right) &= 2.790\times{10^{-2}}-1.780\times{10^{-2}}Z+77.05Z^2\\
\delta_2\left(Z\right) &= 6.730\times{10^{-3}}+2.690Z-52.39Z^2.
\end{split}
\end{equation}
For $1.0\times{10^{-3}} \leq Z \leq 4.0\times{10^{-3}}$, we have

\begin{equation}
\begin{split}
B_1\left(Z\right) &= 40.98+3.415\times{10^4}Z-8.064\times{10^{6}}Z^2 \\
K_1\left(Z\right) &= 35.17+1.548\times{10^4}Z-3.759\times{}10^{6}Z^2\\
K_2\left(Z\right) &= 20.36+1.162\times{10^5}Z-2.276\times{}10^{7}Z^2\\
\delta_1\left(Z\right) &= 2.500\times{10^{-2}}-4.346Z+1.340\times{10^3}Z^2\\
\delta_2\left(Z\right) &= 1.750\times{10^{-2}}+11.39Z-2.902\times{10^3}Z^2.
\end{split}
\end{equation}

Finally, for $Z < 1.0\times{10^{-3}}$, the coefficients do not depend on $Z$,
\begin{equation}
\begin{split}
B_1 &= 67.07 \\
K_1 &= 46.89\\
K_2 &= 1.138\times{10^2} \\
\delta_1 &= 2.199\times{10^{-2}}\\
\delta_2 &= 2.602\times{10^{-2}}.
\end{split}
\end{equation}

\end{document}